\documentclass[]{jpsj3}
\def\lsim{\buildrel {\textstyle <}\over {_\sim}}

\title{
Monte Carlo studies of the ordering of the one-dimensional Heisenberg spin glass with long-range power-law interactions
}
\author{Dao Xuan Viet and Hikaru Kawamura}
\inst{Department of Earth and Space Science, Faculty of Science, Osaka University, Toyonaka 560-0043, Japan}

\date{\today}
\abst{

The nature of the ordering of the one-dimensional Heisenberg spin-glass model with a long-range power-law interaction is studied by extensive Monte Carlo simulations, with particular attention to the issue of the spin-chirality decoupling/coupling. Large system sizes up to $L=4096$ are studied. With varying the exponent $\sigma$ describing the power-law interaction, we observe three distinct types of ordering regimes. For smaller $\sigma$, the spin and the chirality order at a common finite temperature with a common correlation-length exponent, exhibiting the standard spin-chirality coupling behavior. For intermediate $\sigma$, the chirality orders at a temperature higher than the spin, exhibiting the spin-chirality decoupling behavior. For larger $\sigma$, both the spin and the chirality order at zero temperature. We construct a phase diagram in the $\sigma$ versus the temperature plane, and discuss implications of the results. Critical properties associated with both the chiral-glass and the spin-glass transitions are also determined.
}

\kword{spin glass, chiral glass, chirality, long-range interaction, frustration}

\begin{document}
\maketitle

\section{Introduction}

 In spite of quite extensive studies for years, the true nature of the ordering of spin-glass (SG) magnets still remains elusive and controversial \cite{review}. Since the magnetic interaction in most of real SG materials is known to be nearly isotropic, they should be described as a first approximation by the isotropic Heisenberg model. Recently, consensus appears among various numerical works that the isotropic Heisenberg SG in three dimensions (3D) exhibits a finite-temperature transition, while the nature of the transition still remains controversial \cite{Kawamura92,HukuKawa00,HukuKawa05,VietKawamura09,Campos06,LeeYoung07,Fernandez09}. 

 It has been suggested in Ref.\cite{Kawamura92} that the model might exhibit an intriguing ``spin-chirality decoupling'' phenomenon, {\it i.e.\/}, the chirality exhibits the glass order at a temperature higher than the standard SG order, $T_{CG} > T_{SG}$ \cite{HukuKawa00,HukuKawa05,VietKawamura09}. Chirality is a multispin variable representing the handedness of the noncollinear or noncoplanar spin structures induced by frustration. Based on such a spin-chirality decoupling picture of the 3D Heisenberg SG, a chirality scenario of experimental SG transition has been advanced \cite{Kawamura92,Kawamura07,Kawamura09}. By contrast, Refs.\cite{Campos06,LeeYoung07,Fernandez09} claim that the 3D Heisenberg SG does not exhibit such a spin-chirality decoupling, only a single transition which is simultaneously SG and chiral-glass (CG).

 To get deeper insight into the behavior in physical dimension $d=3$, it is often useful to study the phenomena by extending the dimensionality to general $d$ dimensions. In the limit of low $d$, the short-range (SR) Heisenberg SG exhibits only a $T=0$ transition in $d=1$. In $d=2$, recent calculations suggest that the vector SG model, either the three-component Heisenberg SG \cite{KawaYone03} or the two-component {\it XY\/} SG \cite{Weigel08}, exhibits a $T=0$ transition accompanied by the spin-chirality decoupling, {\it i.e.\/}, the CG correlation-length exponent $\nu_{CG}$ is greater than the SG correlation-length exponent $\nu_{SG}$. The spin-chirality decoupling associated with a finite-temperature transition could occur, if any, in $d\geq 3$. In the opposite limit of high $d$, the SR Heisenberg SG model in infinite dimensions $d\rightarrow \infty $ reduces to the mean-field (MF) model, {\it i.e.\/}, the Sherrington-Kirkpatrick (SK) model. The Heisenberg SK model is known to exhibit only a single finite-temperature SG transition, with no spin-chirality decoupling. In high but finite $d$, Monte Carlo (MC) result of Ref.\cite{ImaKawa03} suggested that the spin-chirality decoupling did not occur in $d=5$, but might occur in $d=4$. Reflecting an intrinsic difficulty in thermalizing large systems in high dimensions, however, the true situation still remains largely unclear.

  In the present paper, we attack the issue of the spin-chirality coupling/decoupling in the Heisenberg SG from a different perspective. Namely, we study a different type of Heisenberg SG model, {\it i.e.\/}, the one-dimensional (1D) Heisenberg SG with a long-range (LR) power-law interaction proportional to $1/r^{\sigma}$ ($r$ is the spin distance). A preliminary account of the simulation was presented in Ref.\cite{VietKawamura1D}. 

 In the limit of sufficiently large $\sigma \rightarrow \infty $, the model reduces to the standard $d=1$ model with a SR interaction. In the opposite limit of $\sigma \rightarrow 0$, the model reduces to an infinite-range model, {\it i.e.\/}, the SK model corresponding to $d=\infty $. (Note that for $\sigma \leq 1/2$, one needs to rescale the interaction strength by an appropriate power of $L$ to make the energy extensive.) Hence, varying $\sigma$ of the 1D LR model might be analogous to varying $d$ in the SR model \cite{BhattYoung86}. Indeed, this correspondence was supported by recent studies by Katzgraber and Young \cite{Katzgraber03,Katzgraber05} and  by Leuzzi {\it et al\/} \cite{Leuzzi99,Leuzzi08} for the Ising SG. Indeed, these authors have suggested more detailed correspondence between  $d$ of the SR model and $\sigma$ of the 1D LR model, {\it e.g.\/}, (i) the upper critical dimension  $d=6$ corresponds to $\sigma=2/3$,  (ii) the lower critical dimension, which lies between $d=2$ and 3, corresponds to $\sigma=1$, and (iii)  $d=3$ corresponds to $\sigma \sim 0.9$.

 Advantages of studying such 1D models might be threefold. First, systems of large linear size $L$, never available in high dimensions, can be studied (up to $L=4096$ in the present calculation). Second, one can continuously change and even fine-tune the parameter $\sigma$ playing the role of effective ``dimensionality'', while it is impossible to continuously change the real dimensionality $d$ in the SR model. Hence, by studying the properties of the 1D model with varying $\sigma$, one might get  an overall picture concerning how the ``coupling vs. decoupling'' behavior depends on the effective dimensionality. Third, certain analytical results based on the renormalization-group (RG) calculations are available in 1D, which might be utilized in interpreting the numerical data. 

 Indeed, RG calculations, though did not take account of the possibility of the spin-chirality decoupling, suggested that the model exhibited a rich ordering behavior with varying $\sigma$ \cite{Kotliar82,Chang84}. For $\sigma \leq 2/3$, the Gaussian fixed point is stable and the model exhibited a finite-temperature SG transition of the MF type. For $2/3< \sigma < 1$,  a non-trivial LR fixed point becomes stable leading to a finite-temperature SG transition characterized by the non-MF exponents. In particular, the critical-point-decay exponent is determined solely by the power describing the spin-spin interaction, and is given by $\eta_{SG}=3- 2\sigma$ \cite{Kotliar82,Chang84}. For $\sigma \geq 1$, the SG transition occurs only at zero-temperature with $\eta_{SG}=1$. 

 Meanwhile, it remains to be seen how the spin-chirality decoupling arises in this 1D model with varying $\sigma$. Since the MF Heisenberg SK model does not show the spin-chirality decoupling, the spin-chirality decoupling associated with a finite-temperature transition should be realized, if any, only in the intermediate range of $\sigma$, near or below $\sigma=1$. Thus, we study here both the spin and the chiral orderings of the model by large-scale MC simulations, varying $\sigma$ in the range $0.7\leq \sigma \leq 1.1$, which spans the non-MF regime.  By studying such a one-dimensional Heisenberg SG model with LR power-law interactions, we are able to study the spin-chirality decoupling/coupling phenomena from a wider perspective.  Our numerical results indicate that the model exhibits the spin-chirality decoupling in the range $0.8 \lsim \sigma \lsim 1.1$, while the usual spin-chirality coupling behavior occurs for $\sigma \lsim 0.8$.
 
 The paper is organized as follows. In \S 2, we define our model and explain some of the details of our numerical method employed. Various physical quantities calculated in our simulations are introduced in \S 3. Our MC results are presented in \S 4. Quantities like the CG and SG correlation-length ratios, the CG and SG susceptibilities, the CG and SG Binder ratios, the CG and SG overlap distribution functions, {\it etc\/}, are calculated for various values of the range parameter $\sigma$. Then, a phase diagram of the 1D LR model is constructed in the $\sigma$ versus the temperature plane. Critical properties associated with the CG and SG transitions are analyzed in \S 5 my means of a finite-size scaling analysis. Finally, \S 6 is devoted to summary and discussion.

\section{The model and the method}

 We study the 1D classical Heisenberg model with the random LR power-law interaction $J_{ij}$, whose Hamiltonian is given by
\begin{equation}
{\cal H}=-\sum_{<ij>}J_{ij}\vec{S}_i\cdot \vec{S}_j\ \ ,
%\label{eqn:hamil}
\end{equation}
where $\vec{S}_i=(S_i^x,S_i^y,S_i^z)$ is a three-component unit vector at the $i$-th site, and the $<ij>$ sum is taken over all spin pairs on the chain once. The coupling $J_{ij}$ decays with a geometric distance $r_{ij}$ as a power-law, 
\begin{equation}
J_{ij}=C\frac {\epsilon_{ij}}{r_{ij}^{\sigma }},\ \ \ C=\surd {\frac {L}{\sum_{<ij>} r_{ij}^{-2\sigma}}},
\end{equation}
where $\epsilon_{ij}$ is an independent random Gaussian variable with zero mean and standard deviation unity. Periodic boundary condition is applied by placing $L$ spins on a ring. Then, the geometric distance between the spins at $i$ and $j$ is given by $r_{ij}=(L/{\pi })\sin(\pi \left| i-j \right|/L)$.

 We perform extensive MC simulations for various values of $\sigma=0.7$,
 $0.8$, $0.85$, $0.9$, $0.95$, $1.0$ and $1.1$. Preliminary result for $\sigma=1.1$ was reported in Ref.\cite{MatsuKawa07}. The lattice sizes studied are $L$= 128, 256, 512, 1024, 2048 and 4096. In our simulation, we use a single-spin-flip heat-bath and an over-relaxation method combined with temperature-exchange technique \cite{Hukushima96}. We perform over-relaxation sweeps 5 times per every heat-bath sweep, which constitutes our unit MC step. 

 Equilibration is checked by monitoring: i) All the ``replicas'' travel back and forth many times (typically more than 10 times) along the temperature axis during the temperature-exchange process between maximum and minimum temperature points, whereas the relaxation due to single-spin flip is fast enough (both chiral and spin autocorrelation times about 20 MC steps or less) at the maximum temperature: (ii) All the measured quantities converge to stable values. Error bars of physical quantities are estimated by the sample-to-sample statistical fluctuation over bond realizations.

 In Tables I and II, we show the details of our simulation conditions, including the system size $L$, the number of independent samples (bond realizations) $N_{s}$, the number of temperature points used in the temperature-exchange process $N_{T}$, and the minimum and maximum temperatures $T_{min}$ and $T_{max}$. In Table I, we show these conditions for $\sigma=0.7$, 0.8, 0.85, 0.9, and 0.95 which correspond to the $T_{SG}>0$ regime, and those for $\sigma=1.0$ and 1.1 in Table II which correspond to the $T_{SG}=0$ regime.

\begin{center}
 \begin{table}
  \begin{tabular*} {0.5\textwidth} {@{\extracolsep{\fill}} c c c c c c}
   \hline
   \hline
   $\sigma $ & $L$ & $N_{s}$ & $N_{T}$ & $T_{max}$ & $T_{min}$ \\
   \hline
  0.7 & 128 & 896 & 16 & 0.275 & 0.120 \\
      & 256 & 896 & 16 & 0.275 & 0.120 \\
      & 512 & 896 & 16 & 0.275 & 0.120 \\
      & 1024 & 512 & 16 & 0.275 & 0.120 \\
      & 2048 & 256 & 16 & 0.275 & 0.120 \\
   \hline
  0.8 & 128 & 896 & 16 & 0.190 & 0.090 \\
      & 256 & 896 & 16 & 0.190 & 0.090 \\
      & 512 & 896 & 16 & 0.190 & 0.090 \\
      & 1024 & 512 & 16 & 0.190 & 0.090 \\
      & 2048 & 256 & 16 & 0.190 & 0.090 \\
      & 4096 & 256 &  8 & 0.190 & 0.134 \\
   \hline
  0.85 & 128 & 896 & 20 & 0.170 & 0.065 \\
       & 256 & 896 & 20 & 0.170 & 0.065 \\
       & 512 & 896 & 20 & 0.170 & 0.065 \\
       & 1024 & 512 & 20 & 0.170 & 0.065 \\
       & 2048 & 512 & 10 & 0.170 & 0.108 \\
       & 4096 & 256 & 8 & 0.170 & 0.119 \\
   \hline
  0.9 & 128 & 896 & 20 & 0.150 & 0.055 \\
      & 256 & 896 & 20 & 0.150 & 0.055 \\
      & 512 & 896 & 20 & 0.150 & 0.055 \\
      & 1024 & 896 & 20 & 0.150 & 0.055 \\
      & 2048 & 896 & 13 & 0.135 & 0.072 \\
      & 4096 & 256 & 8 & 0.135 & 0.093 \\
   \hline
  0.95 & 128 & 896 & 16 & 0.113 & 0.050 \\
       & 256 & 896 & 16 & 0.113 & 0.050 \\
       & 512 & 896 & 16 & 0.113 & 0.050 \\
       & 1024 & 896 & 16 & 0.113 & 0.050 \\
       & 2048 & 512 & 16 & 0.113 & 0.050 \\
       & 4096& 256 & 12 & 0.113 & 0.062 \\

   \hline
   \hline
\end{tabular*}
\caption{
Various parameters of our Monte Carlo simulations  of the 1D LR model with $\sigma=0.7$, 0.8, 0.85, 0.9 and 0.95 which correspond to the $T_{SG}>0$ regime. $L$ is the linear dimension of the system, $N_{s}$ is the number of samples, $T_{max}$ and $T_{min}$ are the highest and the lowest temperatures used in the temperature-exchange run, and $N_{T}$ is the total number of temperature points.
}
%\label{table3}
\end{table}
\end{center}
\begin{center}
 \begin{table}
  \begin{tabular*} {0.5\textwidth} {@{\extracolsep{\fill}} c c c c c c}
   \hline
   \hline
   $\sigma $ & $L$ & $N_{s}$ & $N_{T}$ & $T_{max}$ & $T_{min}$ \\ 
   \hline
  1.0  & 128 & 896 & 16 & 0.097 & 0.0337 \\
       & 256 & 896 & 16 & 0.097 & 0.0337 \\
       & 512 & 896 & 16 & 0.097 & 0.0337 \\
       & 1024 & 896 & 16 & 0.097 & 0.0337 \\
       & 2048 & 256 & 32 & 0.101 & 0.0337 \\
   \hline
  1.1  & 128 & 512 & 32 & 1.0 & 0.014 \\
       & 256 & 1024 & 32 & 1.0 & 0.014 \\
       & 512 & 1024 & 16 & 0.079 & 0.014 \\
       & 1024 & 448 & 64 & 0.22 & 0.019 \\
       & 2048 & 256 & 56 & 0.22 & 0.019 \\
   \hline
   \hline
\end{tabular*}
\caption{
Various parameters of our Monte Carlo simulations of the 1D LR model with $\sigma=1.0$ and 1.1 which correspond to the $T_{SG}=0$ regime. $L$ is the linear dimension of the system, $N_{s}$ is the number of samples, $T_{max}$ and $T_{min}$ are the highest and the lowest temperatures used in the temperature-exchange run, and $N_{T}$ is the total number of temperature points.
}
%\label{table4}
\end{table}
\end{center}

\section{Physical quantities}

 In this section, we define various physical quantities calculated in the following section.

 The local chirality at the $i$-th site $\chi_{i}$ is defined for three neighboring Heisenberg spins by the scalar
\begin{equation}
\chi_{i}=
\vec{S}_{i+1}\cdot
(\vec{S}_i\times\vec{S}_{i-1})\ \ .
\end{equation}

First, we define an `overlap' for the chirality. We prepare at each temperature two independent systems 1 and 2 described by the same Hamiltonian (1) with the same interaction set. We simulate these two replicas 1 and 2 in parallel with using different spin initial conditions and different sequences of random numbers. 

The $k$-dependent chiral overlap, $q_\chi(k)$, is defined as an overlap variable between the two replicas 1 and 2 as a scalar
\begin{equation}
q_\chi(k) =
\frac{1}{3N}\sum_{i=1}^N \chi_{i}^{(1)}\chi_{i}^{(2)}e^{i k \cdot r_i},
\end{equation}
where the upper suffixes (1) and (2) denote the two replicas of the system, and $r_{i}$ is the distance along the chain (ring) in units of lattice spacing.

The $k$-dependent spin overlap, $q_{\alpha\beta}(k)$, is defined by a {\it tensor\/} variable between the $\alpha$ and $\beta$ components of the Heisenberg spin,
\begin{equation}
q_{\alpha\beta}(k) = 
\frac{1}{N}\sum_{i=1}^N S_{i\alpha}^{(1)}S_{i\beta}^{(2)}e^{i k r_i},
\ \ \ (\alpha,\beta=x,y,z).
\end{equation}

In term of the $k$-dependent overlap, the CG and SG order parameters are defined by the second moment of the overlap at a wavevector $k=0$,
\begin{equation}
q_{CG}^{(2)}=\frac {[\langle | q_{\chi}(0)|^2 \rangle]} {\overline{\chi}^{4}},
\end{equation}
\begin{equation}
q_{SG}^{(2)} = [\langle q_{\rm s}(0)^2\rangle]\ ,
\ \ \  
q_{\rm s}(k)^2 = \sum_{\alpha,\beta=x,y,z} \left| q_{\alpha\beta}(k) \right| ^2.
\end{equation}
The CG order parameter $q_{CG}^{(2)}$ has been normalized here by the mean-square amplitude of the local chirality,
\begin{equation}
\overline{\chi}^{2}=\frac{1}{3N}\sum_i^N[\langle \chi_{i}^2\rangle],
\end{equation}
which remains nonzero only when the spin has a noncoplanar structure locally. The local chirality amplitude depends weakly on the temperature and the lattice size, in contrast to the Heisenberg spin variable whose amplitude is fixed to be unity by definition.

 One often uses the Binder ratios to estimate the critical temperature. The CG and the SG Binder ratios are defined by
\begin{equation}
g_{CG}=
\frac{1}{2}
\left(3-\frac{[\langle q_{\chi}(0)^4\rangle]}
{[\langle q_{\chi}(0)^2\rangle]^2}\right),
\end{equation}
\begin{equation}
g_{SG} = \frac{1}{2}
\left(11 - 9\frac{[\langle q_{\rm s}(0)^4\rangle]}
{[\langle q_{\rm s}(0)^2\rangle]^2}\right).
%\label{eqn:gs_def}
\end{equation}

These quantities are normalized so that, in the thermodynamic limit, they vanish in the high-temperature phase and gives unity in the non-degenerate ordered state. In the present Gaussian coupling model, the ground state is expected to be non-degenerate so that both $g_{CG}$ and $g_{SG}$ should be unity at $T=0$.

 Finite-size correlation length of the 1D LR model is defined by
\begin{equation}
\xi = 
\frac{1}{2\sin(k_\mathrm{m}/2)}
\left( \frac{ [\langle q(0)^2 \rangle] }
{[\langle q(k_\mathrm{m})^2 \rangle] } -1 \right)^{1/(2\sigma -1)},
\end{equation}
where the $q(k)^2$ is defined via eq.(4) for the chirality and by eqs.(5) and (7) for the spin, with $k_\mathrm{m}=\frac{2\pi}{L}$. The reason for the power $1/(2\sigma -1)$ appearing in eq.(11) is that, at long wavelength, we expect a modified Ornstein-Zernike form for the LR model \cite{Mayor02} 

\begin{equation}
q^{(2)}(k) \propto (t+k^{2\sigma -1})^{-1}, 
\end{equation}
where $t$ is a measure of the deviation from the critical point.

The correlation length divided by the system size $\xi/L$, the correlation-length ratio, is a dimensionless quantity. Around the critical temperature $T_c$, this quantity is expected to obey the finite-size scaling form,
\begin{equation}
\frac {\xi}{L}=\tilde X ((T-T_{c})L^{1/\nu})(1+aL^{-\omega}),
\end{equation}
where $\nu$ is the correlation-length exponent, $\tilde X$ a scaling function, $a$ a constant, and $\omega$ the leading correction-to-scaling exponent.

We also calculate the CG and SG susceptibilities defined by
\begin{equation}
\chi_{CG}=Lq_{CG}^{(2)}\ , \ \ \  \chi_{SG}=Lq_{SG}^{(2)}.
\end{equation}

  While the SG and CG susceptibilities are dimensionfull quantities, they can be made dimensionless by dividing them by $L^{2-\eta}$ where $\eta$ is a critical-point-decay exponent. Generally, the exponent $\eta$ is not known in advance, but in the case of the present LR interaction, $\eta_{SG}$ is determined by the power describing the spin-spin interaction and is given by $\eta_{SG}=3- 2\sigma$ \cite{Kotliar82,Chang84}. Thus, one expects the finite-size scaling form to hold around the critical temperature,
\begin{equation}
\frac {\chi} {L^{2-\eta}}=\tilde Y((T-T_{c})L^{1/\nu})(1+a'L^{-\omega}),
\end{equation}
where $\tilde Y$ is a scaling function and $a'$ is a constant.  As $L \rightarrow \infty $, the $\xi/L$ and the $\chi/(L^{2-\eta})$ curves of different $L$ plotted versus the temperature should asymptotically cross at $T = T_{c}$. Unfortunately, exact expression is not known for the corresponding chiral-glass exponent $\eta_{CG}$. 

 The chiral-overlap distribution $P(q_{\chi})$ is defined by 
\begin{equation}
P(q_{\chi}^{'})=[\langle \delta(q_{\chi}^{'}-q_{\chi}(0))\rangle].
\end{equation}
The spin-overlap distribution is defined originally in the tensor space with $3\times3=9$ components. To make this quantity more easily visible, one may define the diagonal spin-overlap, which is a trace of the original tensor overlap as \cite {HukuKawa05,ImaKawa03} 
\begin{equation}
P(q_{diag})=[\langle \delta(q_{diag}-\sum_{\mu=x,y,z}q_{\mu \mu}(0))\rangle].
\end{equation}

\section{Monte Carlo results}

 In this section, we present the results of our MC simulation on the 1D LR Heisenberg SG for various values of the range parameter $\sigma$.

\subsection{$\sigma=0.90$}

\begin{figure}[!hbp]
\begin{center}
\includegraphics[scale=0.9]{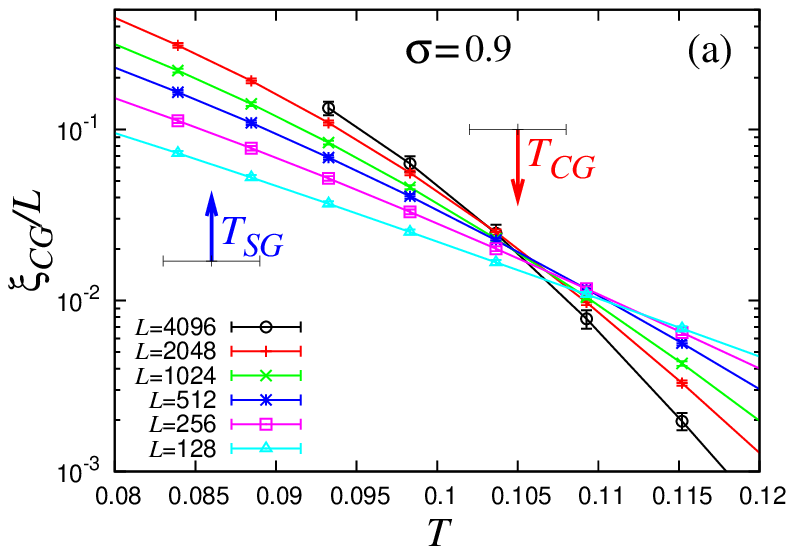}
\includegraphics[scale=0.9]{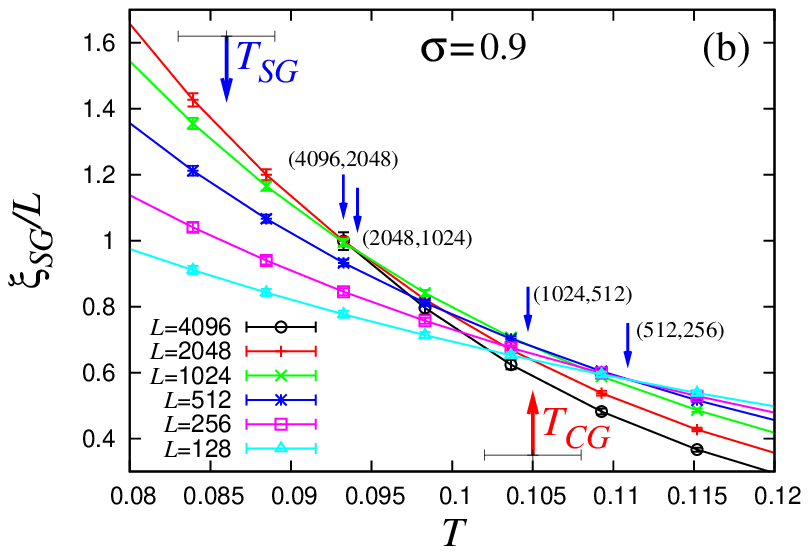}
\end{center}
\caption{
The correlation-length ratio for the chirality (a) and for the spin (b) plotted versus the temperature for $\sigma=0.9$. The red (blue) arrow indicates the bulk chiral-glass (spin-glass) transition point. Note that the $\xi_{CG}/L$ data are given on a semi-logarithmic plot.
} 
\end{figure}
\begin{figure}[!hbp]
\begin{center}
\includegraphics[scale=0.9]{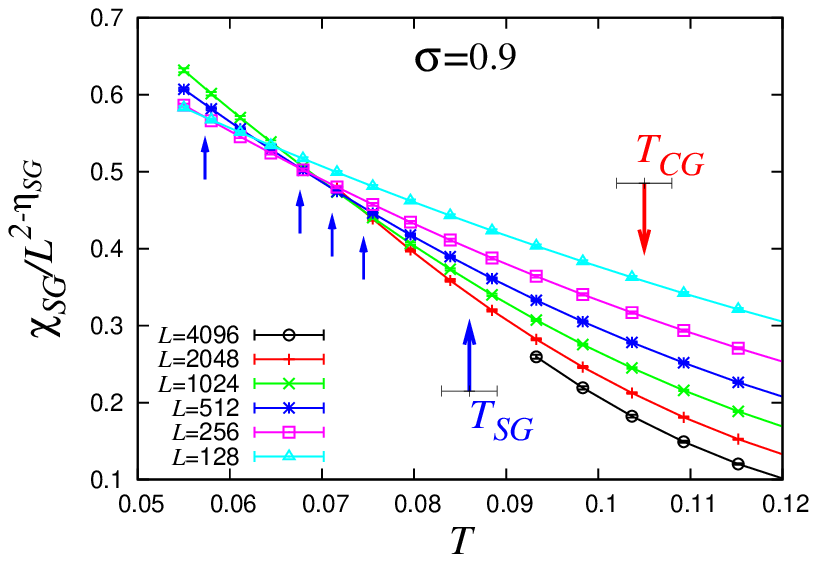}
\end{center}
\caption{
The spin-glass susceptibility ratio $\chi_{SG}/L^{2-\eta_{SG}}$ plotted versus the temperature for $\sigma=0.9$, with an ``exact'' exponent value $\eta_{SG}=3-2\sigma=1.2$. The crossing points are indicated by small blue arrows. The red (blue) arrow indicates the bulk chiral-glass (spin-glass) transition point.
} 
\end{figure}
\begin{figure}[!hbp]
\begin{center}
\includegraphics[scale=0.9]{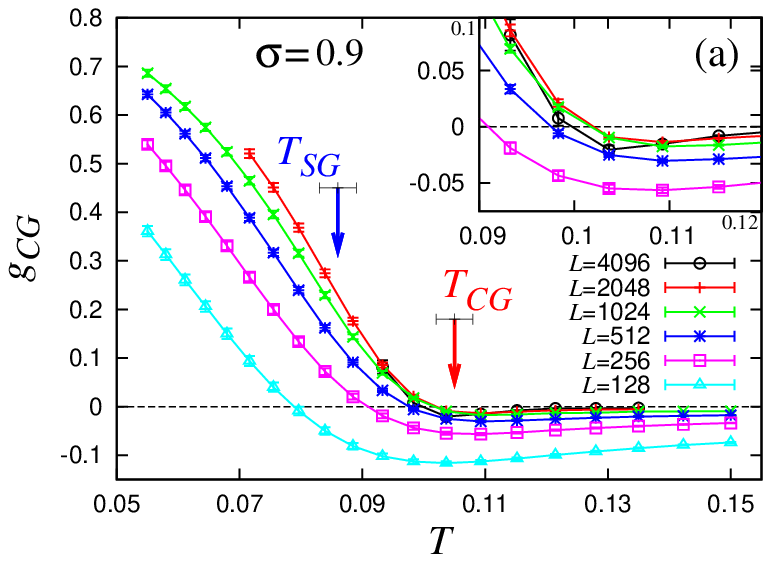}
\includegraphics[scale=0.9]{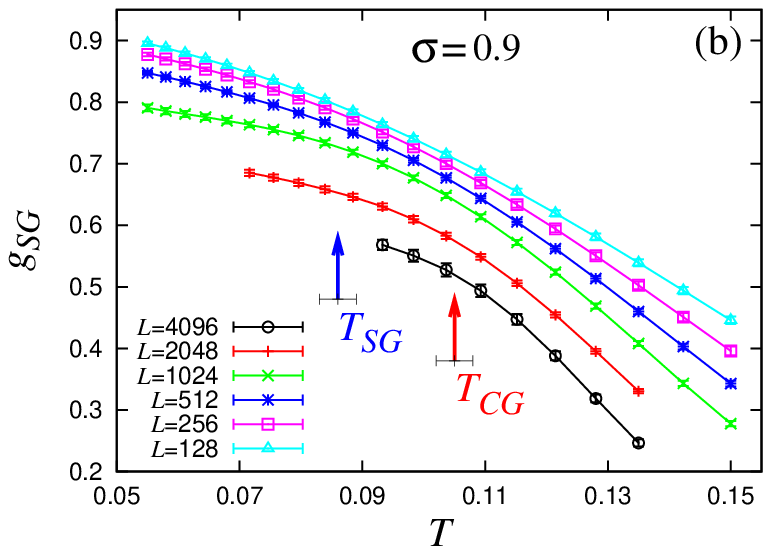}
\end{center}
\caption{
The Binder ratio for the chirality (a) and for the spin (b) plotted versus the temperature for $\sigma=0.9$. The red (blue) arrow indicates the bulk chiral-glass (spin-glass) transition point.
}
\end{figure}
\begin{figure}[!hbp]
\begin{center}
\includegraphics[scale=0.9]{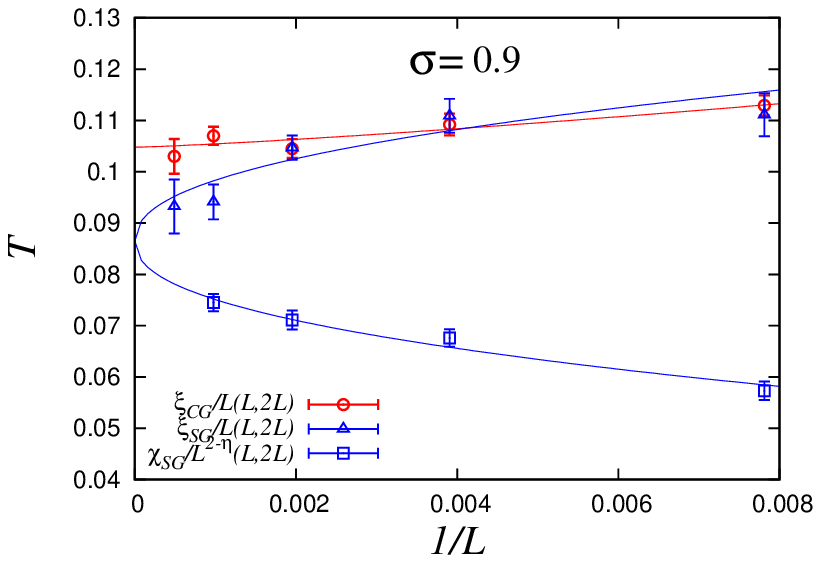}
\end{center}
\caption{
The (inverse) size dependence of the crossing temperatures of $\xi_{CG}/L$, $\xi_{SG}/L$, and $\chi_{SG}/L^{2-\eta_{SG}}$ for $\sigma=0.9$.  Lines represent power-law fits of the form (18). The CG and SG transition temperatures are extrapolated to be $T_{CG}=0.105\pm 0.003$ and $T_{SG}=0.086\pm 0.003$.
} 
\end{figure}
\begin{figure}[!hbp]
\begin{center}
\includegraphics[scale=0.9]{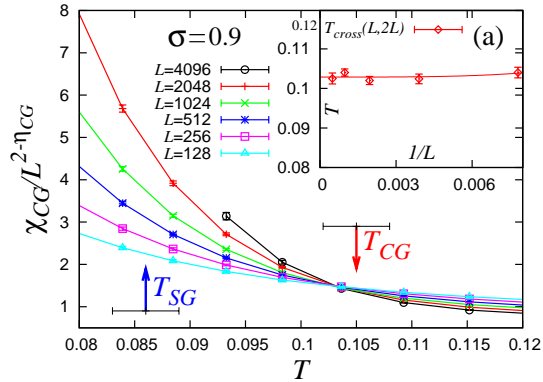}
\end{center}
\caption{
The chiral-glass susceptibility ratio $\chi_{CG}/L^{2-\eta_{CG}}$ plotted versus the temperature for $\sigma=0.9$, where the value of $\eta_{CG}$ is set to $1.9$ as determined by the finite-size-scaling analysis of \S 5. The red (blue) arrow indicates the bulk chiral-glass (spin-glass) transition point. The inset exhibits the (inverse) size dependence of the crossing temperatures in which the line represents a power-law fit of the form (18). The CG transition temperature is extrapolated to be $T_{CG}=0.103\pm 0.003$
} 
\end{figure}
\begin{figure}[!hbp]
\begin{center}
\includegraphics[scale=0.9]{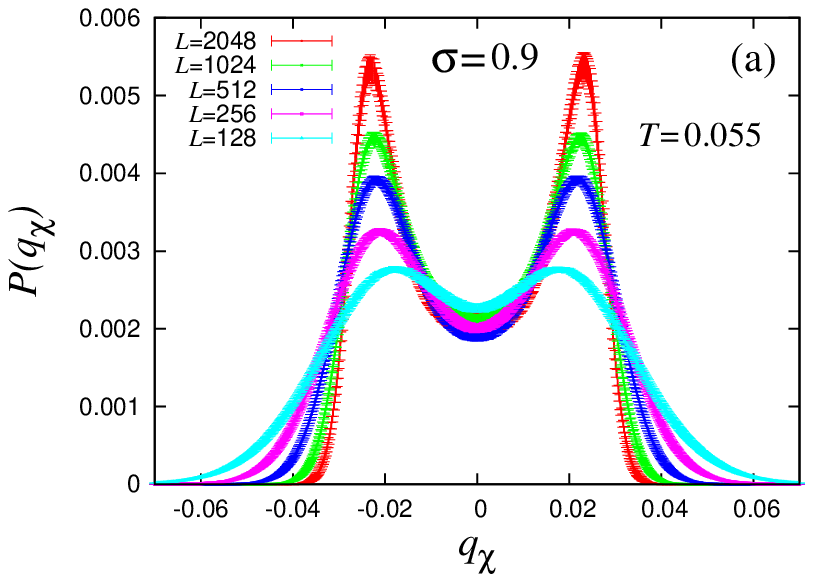}
\includegraphics[scale=0.9]{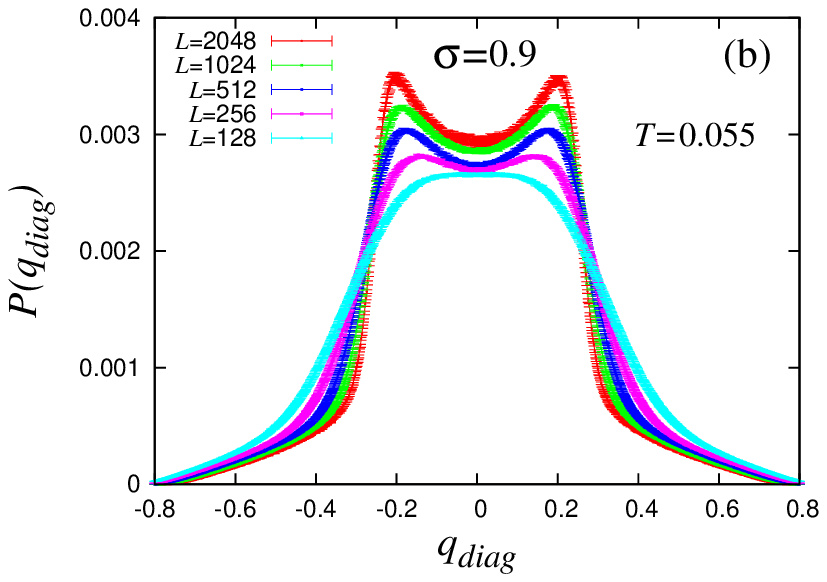}
\end{center}
\caption{
Overlap distribution function of the chirality (a) and of the spin (b) for $\sigma=0.9$ at a temperature $T=0.055$ below $T_{SG}$. The sample average of $L=2048$ is taken here for a subset of total samples (256 samples)
}
\end{figure}

 We begin with the case of $\sigma=0.90$, at which $\sigma$ the spin-chirality decoupling is observed most clearly. In Fig.1, we show the correlation-length ratios for the chirality $\xi_{CG}/L$ (a), and for the spin $\xi_{SG}/L$ (b). As can be seen from the figure, the spin $\xi_{SG}/L$ curves cross at progressively lower temperature as $L$ increases, whereas the chiral $\xi_{CG}/L$ curves intersect at an almost $L$-independent temperature. 
%The crossing temperature $T_{cross}(L)$ of $\xi_{CG}/L$ and of $\xi_{SG}/L$ are plotted in Fig.2 as a function of $1/L_{av}$ for pairs of successive $L$ values, where $L_{av}$ is a mean of the two sizes. 

 We plot in Fig.2 the temperature dependence of the SG susceptibility ratio $\chi_{SG}/L^{2-\eta_{SG}}$ where $2-\eta_{SG}=2\sigma-1=0.8$ for the present value of $\sigma=0.9$. As can be seen from the figure, the data of different $L$ exhibit a crossing behavior as expected for the dimensionless quantity with a finite-$T$ transition. While the crossing points of $\xi_{SG}/L$ approach the bulk SG transition point from above, those of $\chi_{SG}/L^{2-\eta_{SG}}$ approach the SG transition point from below. For the CG susceptibility ratio, on the other hand, this type of analysis has only restricted utility because of the lack of our knowledge of the chirality-chirality interaction and the associated $\eta_{CG}$-value.

 The Binder ratios are shown in Fig.3 for the chirality (a), and for the spin (b). The chiral Binder ratio $g_{CG}$ exhibits a shallow negative dip, which tends to be even shallower with increasing $L$, in contrast to the behavior observed for $g_{CG}$ of the 3D SR Heisenberg SG where $g_{CG}$ of different sizes exhibited a crossing behavior on the negative side of $g_{CG}$ \cite{VietKawamura09}. In contrast to this, $g_{CG}$ of the present 1D model hardly exhibits a clear crossing behavior for smaller lattices. Interestingly, however, large lattices of $L=2048$ and $4096$ eventually exhibit two crossings, one at $T\simeq 0.11$ on the negative side of $g_{CG}$ and the other at $T\simeq 0.09$ on the positive side of $g_{SG}$: See the inset of Fig.3(a).

 In order to estimate the bulk CG and SG transition temperatures quantitatively, we need to extrapolate the crossing temperatures of either the correlation-length ratio or the SG susceptibility ratio to $L=\infty$. In Fig.4, we plot $T_{cross}(L)$ of the $\xi_{CG}/L$ curves between of the sizes $L$ and $2L$  as a function $1/L$, together with the corresponding ones of the $\xi_{SG}/L$ and the $\chi_{SG}/L^{2-\eta_{SG}}$ curves.  The error bar of each $T_{cross}(L)$ is estimated on the basis of the bootstrap method over available samples, combined with a polynomial fit (of the fourth or the fifth order) of the temperature dependence of the physical quantities. 
 
 An extrapolation to $L=\infty$ is made on the basis of the relation, 
\begin{equation}
T_{cross}(L)-T_{cross}(\infty)\approx c L^{-\theta}, 
\end{equation}
where $c$ is a constant.  For both cases of the SG and the CG, the exponent $\theta$ is equal to $\frac{1}{\nu}+\omega$ for the crossing temperature $T_{cross}(L)$ of either the correlation-length ratio or the glass-susceptibility ratio.

 In the case of the SG, we perform a combined fit of both $\xi_{SG}/L$ and of $\chi_{SG}/L^{2-\eta_{SG}}$, where a common $T_{cross}(\infty)=T_{SG}$ and $\theta_{SG}$ are assumed. We then get $T_{SG}=0.086\pm 0.003$ and $\theta_{SG}=0.44 \pm 0.07$: See Fig.4. The errors shown here and below are obtained via the standard $\chi^2$-analysis as explained in some detail in Ref.\cite{VietKawamura09}. The $\chi^2$-value per degree of freedom of our fit here is $\chi^2$/DOF $\simeq 1.24$, with the associated fitting probability $Q \simeq 0.29$, which is quite reasonable. The smallness of the error bar associated with $T_{SG}$ comes from the fact that a combined fit of the two independent of $T_{cross}(L)$, {\it i.e.\/}, those of $\xi_{SG}/L$ and of $\chi_{SG}/L^{2-\eta_{SG}}$, is used here, each approaching $T_{SG}$ either from above or from below.  For the CG, we have $T_{cross}(L)$ of $\xi_{CG}/L$ only, since $\eta_{CG}$ is not known in advance. The CG transition temperature is then estimated via a power-law fit of $T_{cross}(L)$ of $\xi_{CG}/L$ to be $T_{CG}=0.105 \pm 0.003$ with the associated $\theta_{CG}= 1.2\pm 1.4$ ($\chi^2$/DOF $\simeq 1.14$ with $Q \simeq 0.32$). The smallness of the error bar associated with $T_{CG}$ comes from the fact that $T_{cross}(L)$ of $\xi_{CG}/L$ exhibits a nearly $L$-independent behavior. Large error bar associated with $\theta_{CG}$ is merely a consequence of this near $L$-independence of $T_{cross}(L)$. Hence, the estimated CG transition temperature, $T_{CG}=0.105 \pm 0.003$, turns out to be higher than the SG transition temperature, $T_{SG}=0.086\pm 0.003$, by about $20 \%$, suggesting that the spin-chirality decoupling certainly occurs at $\sigma=0.90$.

 One may wonder if the CG susceptibility ratio might be utilized in some way in estimating $T_{CG}$. As mentioned above, no exact knowledge is available for the value of $\eta_{CG}$. Yet, a finite-size scaling analysis performed later in \S 5 gives an estimate of $\eta_{CG}\simeq 1.9$ at $\sigma=0.9$. Thus, we show in Fig.5 the temperature dependence of the CG susceptibility ratio $\chi_{CG}/L^{2-\eta_{CG}}$ with $\eta_{CG}=1.9$. The data exhibit a clear crossing at an almost $L$-independent temperature. A power-law extrapolation to $L=\infty$ yields an estimate of $T_{CG}=0.103\pm 0.003$: See the inset. This estimate agrees with the estimate above obtained from $\xi_{CG}/L$, $T_{CG}=0.105\pm 0.003$. 

 As can be seen from Fig.3(b), the spin Binder ratio $g_{SG}$ for larger lattices rapidly decreases with increasing $L$, though a near-merging behavior is observed for smaller lattices at a temperature $T \simeq 0.1$ close to the CG transition temperature. It should be noticed that, at the SG transition temperature, the spin Binder ratio $g_{SG}$ for larger lattices does not exhibit any crossing in sharp contrast to the standard crossing behavior observed in the MF SK model or the 3D Ising SG, but instead, exhibits only a weak wavy structure somewhat similar to the one observed in the 3D SR Heisenberg SG \cite{VietKawamura09}. Presumably, this weak structure would further develop into a nontrivial behavior at $T=T_{SG}$ for still larger lattices. Unfortunately, we cannot tell its detailed form at the present stage.

 In Figs.6(a) and (b), we show the chiral-overlap distribution function (a), and the diagonal-spin-overlap distribution function (b) at a temperature $T=0.055$ which lies well below $T_{CG}$ and $T_{SG}$. The chiral $P(q_{\chi})$ exhibits double peaks at $q_{\chi}=\pm q_{\chi}^{EA}$, which tend to diverge with increasing $L$. Unlike the behavior observed in the 3D Heisenberg SG model \cite{VietKawamura09}, no central peak at $q_{\chi}=0$ is observed for any size studied. Indeed, with increasing $L$, the value of $P(q_{\chi}=0)$ gradually decreases for smaller lattices, but appears to approach a nonzero value of $\simeq 0.002$ for larger lattices. Such features of $P(q_{\chi})$ are different from the features of the 3D SR Heisenberg SG, which exhibits a distinct central peak possibly associated with a one-step-like RSB \cite{VietKawamura09}. Meanwhile, since the behavior of the Binder ratio of the present model has turned out to be entirely different from that of a full RSB \cite{ImaKawa03}, the ordered state of the present model should not simply be regarded as being similar to the one of the MF SK model or of the 3D Ising SG model.

 The diagonal-spin-overlap distribution $P(q_{diag})$ at a temperature $T=0.055$ below $T_{SG}$ shown in Fig.6(b) also exhibits double peaks located at $q_{diag}\simeq \pm 0.2$, which tend to diverge with increasing $L$. The observed diverging peak is the one expected in the SG ordered state of the isotropic Heisenberg SG to arise at $q_{diag}=\pm \frac{1}{3} q^{EA}$ \cite{ImaKawa03}. Hence, our data of $P(q_{diag})$ are consistent with a finite SG LR order occurring at this temperature.

Thus, we have fairly strong numerical evidence of the occurrence of the spin-chirality decoupling for $\sigma=0.90$, {\it i.e.\/}, $T_{CG}=0.105 \pm0.003$ and $T_{SG}=0.086 \pm0.003$.

\subsection{$\sigma=0.85$ {\rm and} $\sigma=0.95$}

Next, we investigate the cases of $\sigma=0.85$ and $0.95$, where the spin-chirality decoupling is also likely to occur. The behaviors of the correlation-length ratio and the Binder ratio are qualitatively similar to those for $\sigma=0.90$ shown above. Hence, we omit exhibiting the corresponding figures just to save space. Meanwhile, a qualitative change occurs in the behavior of the SG susceptibility ratio for $\sigma=0.95$, where the crossing is no longer observed in the temperature and the lattice-size range investigated in contrast to the crossing behavior shown in Fig.2. This is demonstrated in Fig.7.

\begin{figure}[!hbp]
\begin{center}
\includegraphics[scale=0.9]{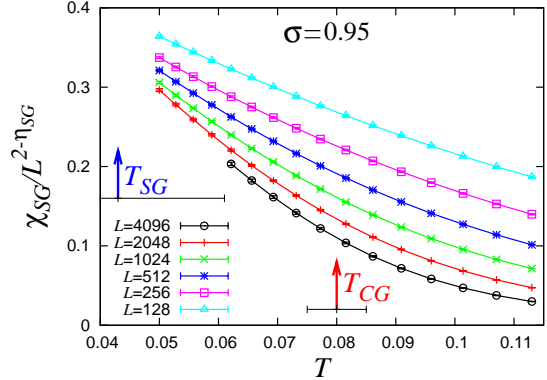}
\end{center}
\caption{
The spin-glass susceptibility ratio $\chi_{SG}/L^{2-\eta_{SG}}$ versus the temperature for $\sigma=0.95$, with an ``exact'' exponent value $\eta_{SG}=3-2\sigma=1.1$. 
} 
\end{figure}
\begin{figure}[!hbp]
\begin{center}
\includegraphics[scale=0.9]{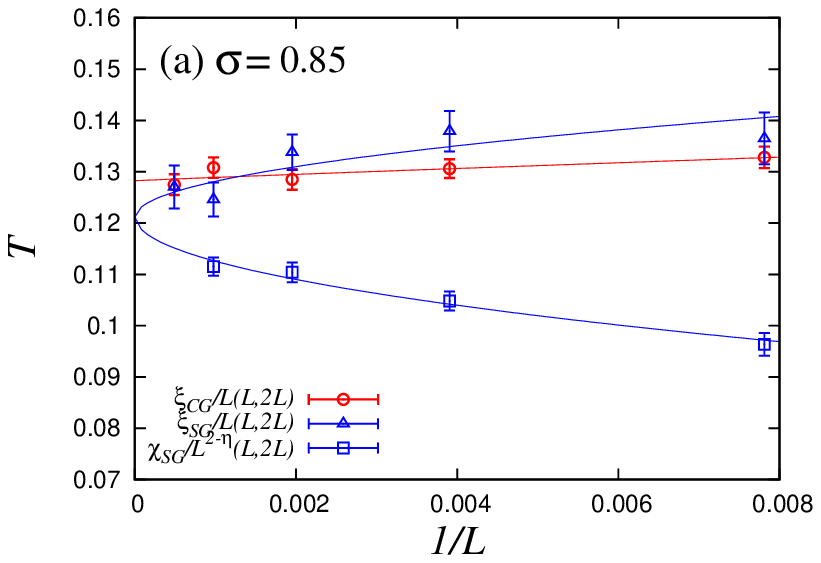}
\includegraphics[scale=0.9]{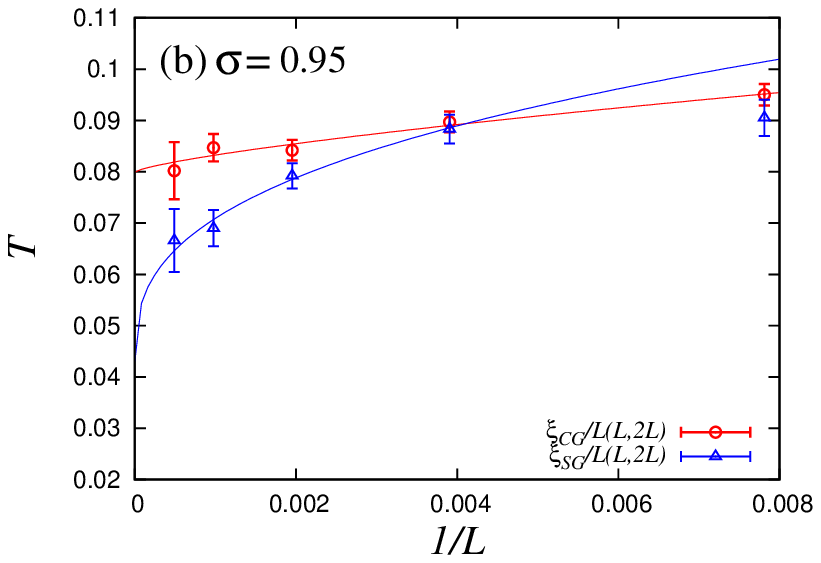}
\end{center}
\caption{
The (inverse) size dependence of the crossing temperatures of $\xi_{CG}/L$, $\xi_{SG}/L$, and $\chi_{SG}/L^{2-\eta_{SG}}$ for $\sigma=0.85$ (a) and for $\sigma=0.95$ (b).  Lines represent power-law fits of the form (18). The CG and SG transition temperatures are extrapolated to be $T_{CG}=0.128\pm 0.003$ and $T_{SG}=0.121\pm 0.003$ for $\sigma=0.85$, and  $T_{CG}=0.080\pm 0.005$, and $T_{SG}=0.043^{+0.018}_{-0.042}$ for $\sigma=0.95$.
} 
\end{figure}

 We show in Fig.8(a) the size dependence of the crossing temperatures $T_{cross}(L)$ of $\xi_{CG}/L$, $\xi_{SG}/L$ and $\chi_{SG}/L^{2-\eta_{SG}}$, for the case of $\sigma=0.85$. The SG transition temperature can be estimated from $T_{cross}(L)$ of $\xi_{SG}/L$ and of $\chi_{SG}/L^{2-\eta_{SG}}$ fairly accurately, each approaching $T_{SG}$ either from above or from below. The combined fit of these two quantities based on eq.(18) then yields $T_{SG}=0.121 \pm 0.003$ and $\theta_{SG}=0.50 \pm 0.09$ ($\chi^2$/DOF $\simeq 0.84$ with $Q \simeq 0.52$). The CG transition temperature is estimated by a power-law fit of $T_{cross}(L)$ of $\xi_{CG}/L$ to be $T_{CG}=0.128 \pm 0.003$ with $\theta_{CG}=0.95\pm 1.58$ ($\chi^2$/DOF $\simeq 0.72$ with $Q \simeq 0.49$). Hence, $T_{SG}$ lies slightly below $T_{CG}$ by about 6\%.

 In Fig.8(b), we show the size dependence of the crossing temperatures $T_{cross}(L)$ of $\xi_{CG}/L$ and of $\xi_{SG}/L$, for the case of $\sigma=0.95$.  The SG transition temperature is estimated by a power-law fit of $T_{cross}(L)$ of $\xi_{SG}/L$ to be $T_{SG}=0.043^{+0.018}_{-0.042}$, with the associated $\theta_{SG}=0.36 \pm 0.44$ ($\chi^2$/DOF $\simeq 0.42$ with $Q \simeq 0.52$). Note that, for $\sigma=0.95$, we have only one type of $T_{cross}(L)$, {\it i.e.\/}, that of $\xi_{SG}/L$, which exhibits a pronounced decreasing tendency with respect to $L$, leading to a rather low estimate of $T_{SG}$ with the larger error bar. The CG transition temperature is estimated by a power-law fit of $T_{cross}(L)$ of $\xi_{CG}/L$ to be $T_{CG}=0.080 \pm 0.005$ with the associated $\theta_{CG}=0.74 \pm 0.51$ ($\chi^2$/DOF $\simeq 0.43$ with $Q \simeq 0.65$). Hence, $T_{CG}$ is higher than $T_{SG}$ at $\sigma=0.95$.

\subsection{$\sigma=0.8$}

\begin{figure}[!hbp]
\begin{center}
\includegraphics[scale=0.9]{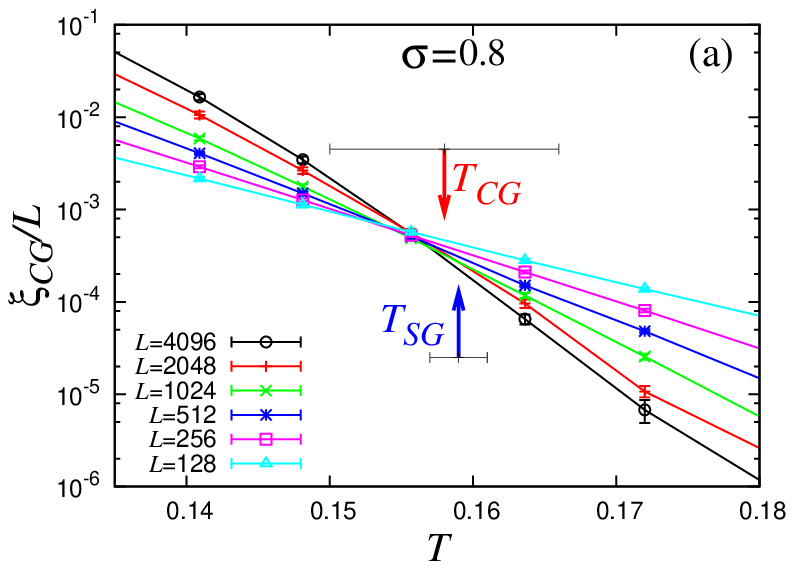}
\includegraphics[scale=0.9]{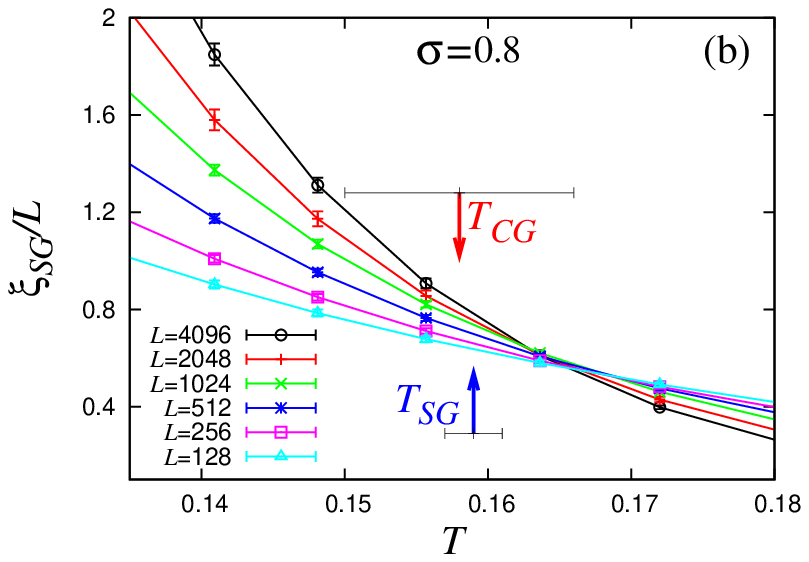}
\end{center}
\caption{
The correlation-length ratio versus the temperature for the chirality (a), and for the spin (b), for $\sigma=0.8$. The red (blue) arrow indicates the bulk chiral-glass (spin-glass) transition point. Note that the $\xi_{CG}/L$ data are given on a semi-logarithmic plot.
} 
\end{figure}
\begin{figure}[!hbp]
\begin{center}
\includegraphics[scale=0.9]{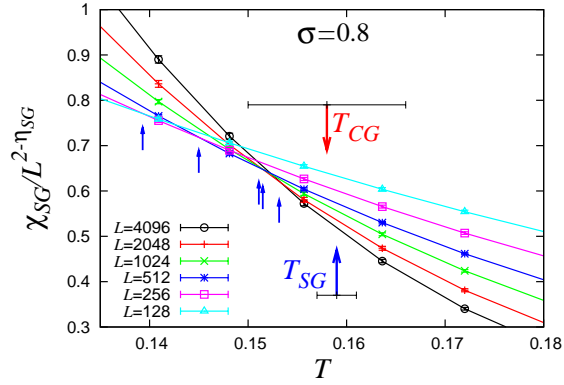}
\end{center}
\caption{
The spin-glass susceptibility ratio $\chi_{SG}/L^{2-\eta_{SG}}$ versus the temperature for $\sigma=0.8$, with an ``exact'' exponent value $\eta_{SG}=3-2\sigma=1.4$. The crossing points are indicated by small blue arrows. The red (blue) arrow indicates the bulk chiral-glass (spin-glass) transition point.
} 
\end{figure}

 Next, we study the case of $\sigma=0.8$, at which the spin-chirality decoupling ceases to occur as we shall see. In Figs.9-11, we show the correlation-length ratios for the chirality (a) and for the spin (b), the SG susceptibility ratio, and the Binder ratios for the chirality (a) and for the spin (b), respectively. As can be seen from Fig.9, at this value of $\sigma$, the crossing temperatures of the spin $\xi_{SG}/L$ come above those of the chiral $\xi_{CG}/L$, the former (the latter) decreases (increases) with increasing $L$. As can be seen from Fig.10, the SG susceptibility ratio $\chi_{SG}/L^{2-\eta_{SG}}$ shows a clear crossing behavior, where $T_{cross}(L)$ increases with $L$. As can be seen from Fig.11, the chiral Binder ratio exhibits a shallow negative dip, while the spin Binder ratio exhibits a near-merging behavior below $T_{SG}$.

\begin{figure}[!hbp]
\begin{center}
\includegraphics[scale=0.9]{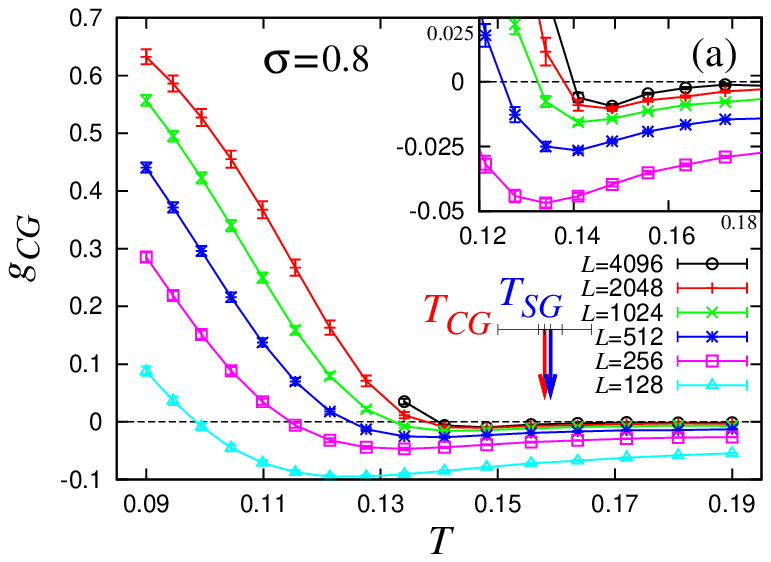}
\includegraphics[scale=0.9]{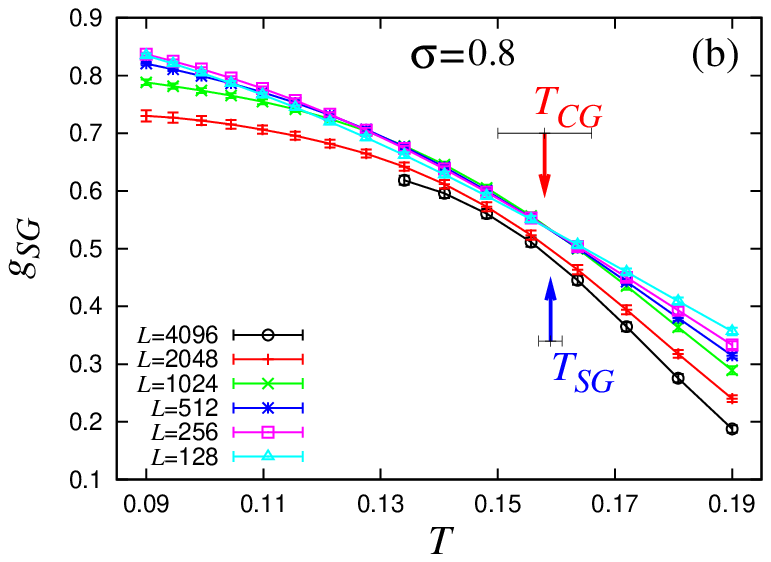}
\end{center}
\caption{
The Binder ratio for the chirality (a) and for the spin (b) plotted versus the temperature for $\sigma=0.8$. The red (blue) arrow indicates the bulk chiral-glass (spin-glass) transition point.
}
\end{figure}

\begin{figure}[!hbp]
\begin{center}
\includegraphics[scale=0.9]{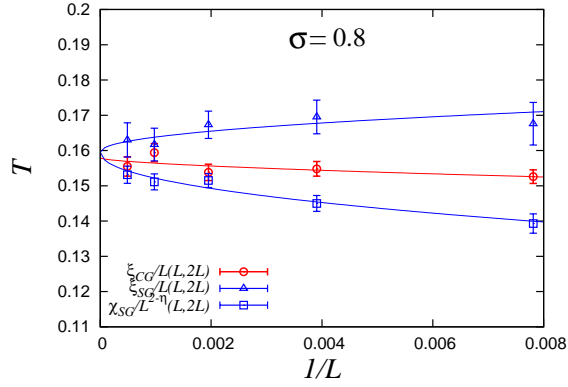}
\end{center}
\caption{
The (inverse) size dependence of the crossing temperatures of $\xi_{CG}/L$, $\xi_{SG}/L$ and $\chi_{SG}/L^{2-\eta_{SG}}$ for $\sigma=0.8$.  Lines represent power-law fits of the form (18). The CG and SG transition temperatures are extrapolated to be $T_{CG}=0.158\pm 0.008$ and $T_{SG}=0.159\pm 0.002$, {\it i.e.\/}, $T_{CG}=T_{SG}$ within the error bar.
} 
\end{figure}

 We show in Fig.12 the size dependence of the crossing temperatures $T_{cross}(L)$ of $\xi_{CG}/L$, $\xi_{SG}/L$ and $\chi_{SG}/L^{2-\eta_{SG}}$ for the case of $\sigma=0.8$. The SG transition temperature is estimated by a combined power-law fit of $T_{cross}(L)$ of $\xi_{SG}/L$ and of $\chi_{SG}/L^{2-\eta_{SG}}$.  Assuming a common $T_{cross}(\infty)=T_{SG}$ and $\theta_{SG}$ in eq.(18), we get $T_{SG}=0.158 \pm 0.008$ with the associated $\theta_{SG}=0.62 \pm 1.95$ ($\chi^2$/DOF $\simeq 0.43$ with $Q \simeq 0.86$). The CG transition temperature is estimated via a power-law fit of $T_{cross}(L)$ of $\xi_{CG}/L$ to be $T_{CG}=0.159 \pm 0.002$ and $\theta_{CG}=0.47 \pm 0.11$ ($\chi^2$/DOF $\simeq 1.14$ with $Q \simeq 0.32$).

\subsection{$\sigma=0.7$}

 Next, we study the case of $\sigma=0.7$, which lies close to the lower critical $\sigma=\frac{2}{3}$, {\it i.e.\/}, the boundary between the non-MF and the MF regimes. The behaviors of the correlation-length ratios, the SG susceptibility ratio, and the Binder ratios turn out to be more or less similar to the ones observed for $\sigma=0.8$. One difference is that the spin Binder ratio $g_{SG}$ now exhibits a rather clear crossing even for smaller lattices on the positive side of $g_{SG}$, as shown in Fig.13.

\begin{figure}[!hbp]
\begin{center}
\includegraphics[scale=0.9]{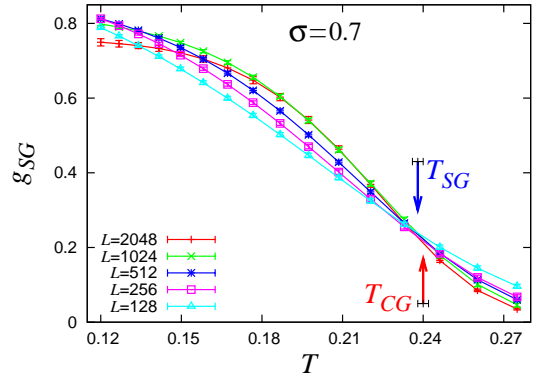}
\end{center}
\caption{
The Binder ratio for the spin plotted versus the temperature for $\sigma=0.7$. The red (blue) arrow indicates the bulk chiral-glass (spin-glass) transition point.
}
\end{figure}

\begin{figure}[!hbp]
\begin{center}
\includegraphics[scale=0.9]{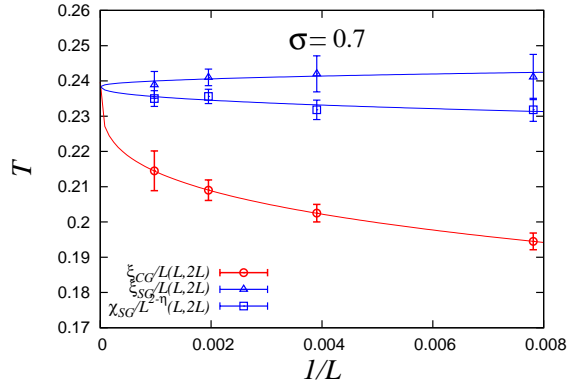}
\end{center}
\caption{
The (inverse) size dependence of the crossing temperatures of $\xi_{CG}/L$, $\xi_{SG}/L$, and $\chi_{SG}/L^{2-\eta_{SG}}$ for $\sigma=0.7$.  Lines represent power-law fits of the form (18). The CG and SG transition temperatures are extrapolated to be $T_{CG}=0.240\pm 0.002$ and $T_{SG}=0.238\pm 0.002$.
} 
\end{figure}

 We show in Fig.14 the size dependence of the crossing temperatures $T_{cross}(L)$ of $\xi_{SG}/L$, $\xi_{CG}/L$, and $\chi_{SG}/L^{2-\eta_{SG}}$. The SG transition temperature is estimated by a combined power-law fit of $T_{cross}(L)$ of $\xi_{SG}/L$ and of $\chi_{SG}/L^{2-\eta_{SG}}$. Assuming a common $T_{cross}(\infty)=T_{SG}$ and $\theta_{SG}$ in eq.(18), we get $T_{SG}=0.238 \pm 0.002$ and $\theta_{SG}=0.44 \pm 0.21$ ($\chi^2$/DOF $\simeq 0.19$ with $Q \simeq 0.94$). The CG transition temperature is estimated via a power-law fit of $T_{cross}(L)$ of $\xi_{CG}/L$ to be $T_{CG}=0.240\pm 0.002$ with the associated $\theta_{CG}=0.28 \pm 0.02$ ($\chi^2$/DOF $\simeq 0.0004$ with $Q \simeq 0.98$). Hence, $T_{SG}$ and $T_{CG}$ agree within the error bar.

\subsection{$\sigma=1.0$}

\begin{figure}[!hbp]
\begin{center}
\includegraphics[scale=0.9]{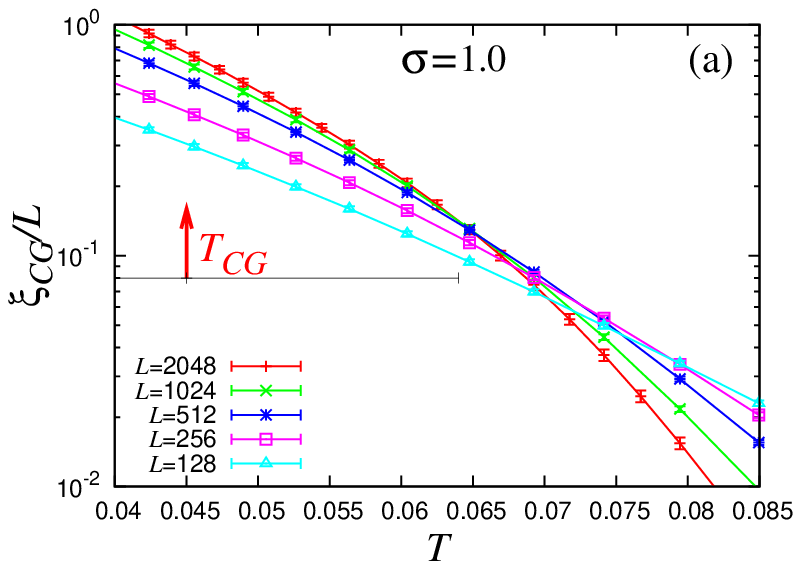}
\includegraphics[scale=0.9]{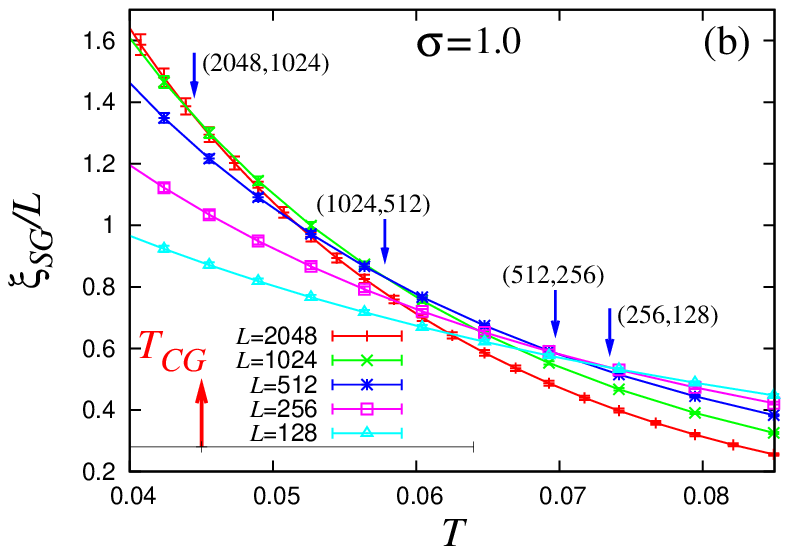}
\end{center}
\caption{
The correlation-length ratio for the chirality (a) and for the spin (b) plotted versus the temperature for $\sigma=1.0$. The red arrow indicates the bulk chiral-glass transition point. Note that the $\xi_{CG}/L$ data are given on a semi-logarithmic plot.
} 
\end{figure}
\begin{figure}[!hbp]
\begin{center}
\includegraphics[scale=0.9]{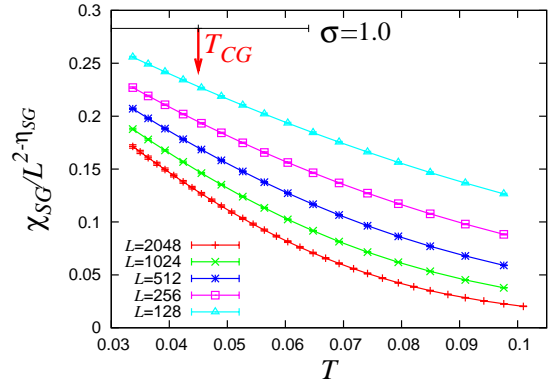}
\end{center}
\caption{
The spin-glass susceptibility ratio $\chi_{SG}/L^{2-\eta_{SG}}$ plotted versus the temperature for $\sigma=1.0$, with an ``exact'' exponent value $\eta_{SG}=3-2\sigma=1$. The red arrow indicates the bulk chiral-glass transition point.
} 
\end{figure}

 Next, we turn to the $T_{SG}=0$ regime, {\it i.e.\/}, the region of $\sigma\geq 1$. We first study the case of $\sigma=1.0$, which is just at the upper critical $\sigma$-value unity, {\it i.e.\/}, at the boundary between the $T_{SG}>0$ regime and the $T_{SG}=0$ regime. In Figs.15-17, we show the correlation-length ratios for the chirality (a) and for the spin (b), the SG susceptibility ratio and the Binder ratios for the chirality (a) and for the spin (b), respectively. As can be seen from Fig.15, the crossing temperatures of both the spin $\xi_{SG}/L$ and of the chiral $\xi_{CG}/L$ curves tend to decrease toward lower temperature as $L$ increases. The SG susceptibility ratio $\chi_{SG}/L^{2-\eta_{SG}}$ is shown in Fig.16. Similarly to the one observed for $\sigma=0.95$, it does not show a crossing in the temperature and the lattice-size range investigated. As can be seen from Fig.17, the chiral Binder ratio exhibits a shallow negative dip. As in the case of $\sigma=0.90$ and $0.95$, the chiral Binder ratio of our two largest sizes $L=1024$ and 2048 exhibits a weak crossing (or a merging) on the positive side of $g_{CG}$ around $T\simeq 0.05$, in addition to the one  around $T\simeq 0.08$ on the negative side of $g_{CG}$.

\begin{figure}[!hbp]
\begin{center}
\includegraphics[scale=0.9]{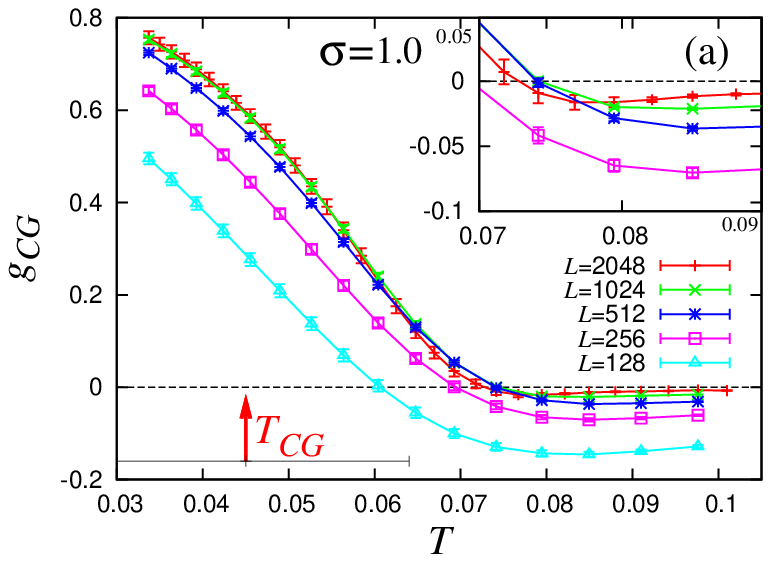}
\includegraphics[scale=0.9]{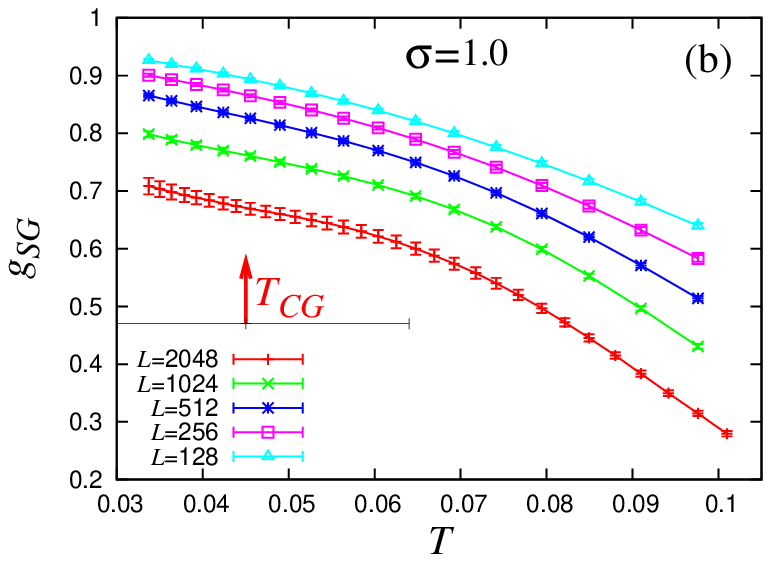}
\end{center}
\caption{
The Binder ratio for the chirality (a) and for the spin (b) plotted versus the temperature for $\sigma=1.0$. The red arrow indicates the bulk chiral-glass transition point.
}
\end{figure}

\begin{figure}[!hbp]
\begin{center}
\includegraphics[scale=0.9]{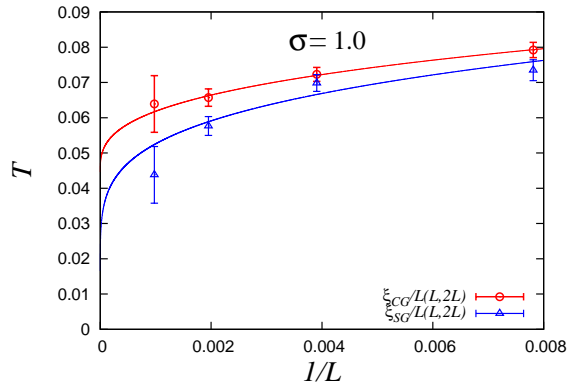}
\end{center}
\caption{
The (inverse) size dependence of the crossing temperatures of $\xi_{CG}/L$, $\xi_{SG}/L$ and $\chi_{SG}/L^{2-\eta_{SG}}$ for $\sigma=1.0$.  Lines for the CG data (red) represent power-law fits of the form (18), while lines for the SG data (blue) represent logarithmic-law fits of the form (19). The chiral-glass transition temperature is extrapolated to be $T_{CG}= 0.045^{+0.019}_{-0.027}$.
} 
\end{figure}

\begin{figure}[ht]
\begin{center}
\includegraphics[scale=0.9]{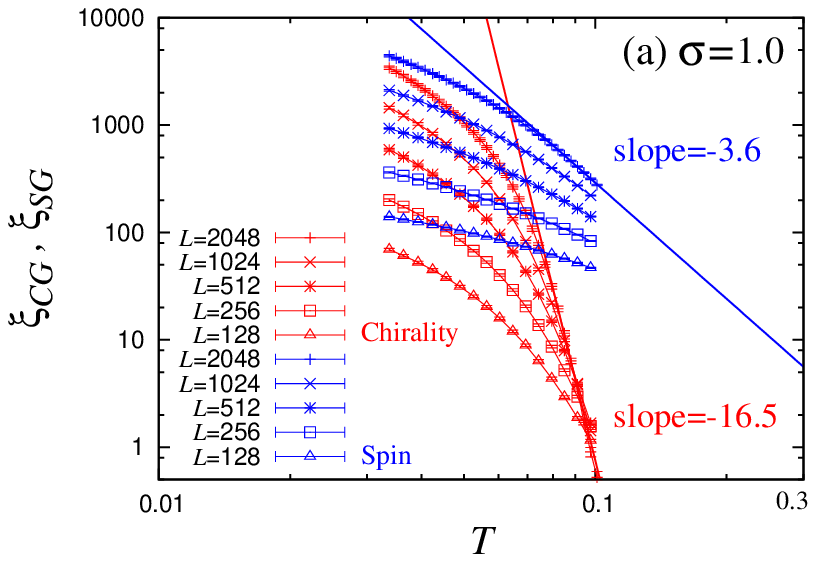}
\includegraphics[scale=0.9]{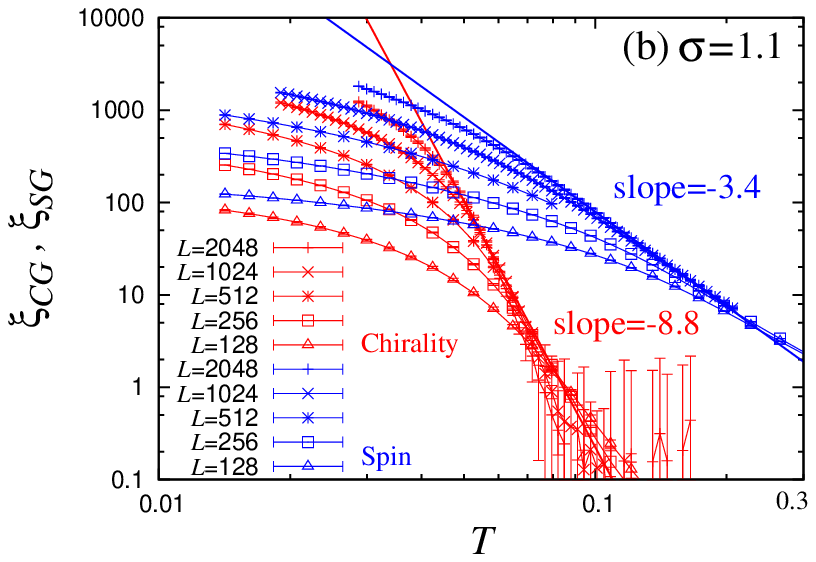}
\end{center}
\caption{
(Color online) The correlation length versus the temperature on a log-log plot for $\sigma=1.0$ (a), and for $\sigma=1.1$ (b).
}
\end{figure}

 We show in Fig.18 the size dependence of the crossing temperatures $T_{cross}(L)$ of $\xi_{CG}/L$ and of $\xi_{SG}/L$.  For the spin, the decreasing tendency of $T_{cross}(L)$ with $L$ becomes pronounced. In fact, a power-law fit becomes unstable here, leading to an indefinitely negative $T_{SG}$-value. Rather, a logarithmic fit of the form expected for the $T=0$ transition at the upper-critical $\sigma$,
\begin{equation}
T_{cross}(L) = b(\ln L + c)^{-\theta}, 
\end{equation}
yields an acceptable fit with $\theta \simeq 2.1$ ($\chi^2$/DOF$=3.88$ and $Q=0.0044$) as shown in Fig.18. This observation supports the  $T=0$ SG transition theoretically expected. For the chirality, a power-law fit of $T_{cross}(L)$ of $\xi_{CG}/L$ yields $T_{CG}=0.045^{+0.019}_{-0.027}$ and $\theta_{CG}=0.34\pm 0.34$ ($\chi^2$/DOF=0.15 and $Q=0.70$).  Hence, $T_{CG}$ is likely to be nonzero at $\sigma=1.0$, although the possibility of $T_{CG}=0$ cannot be ruled out. Indeed, a logarithmic fit of the form (19) also yields an acceptable fit with $\theta \simeq 0.72$ with the $\chi^2$-value comparable to that of the power-law fit. Note, however, even in this case one has $\theta_{CG}\simeq 0.72 << \theta_{SG}\simeq 2.1$, which means that the spin and the chirality are decoupled, {\it i.e.\/}, $\xi_{CG}/\xi_{SG}\rightarrow \infty$. 

 In Fig.19(a), we show the temperature dependence of the SG correlation length of various sizes on a log-log plot together with that of the CG correlation length, which is compared with the ones at $\sigma=1.1$ (b) where one also expects $T_{SG}=0$. The slope of an asymptotic straight line should give an estimate of the exponent $\nu_{SG}$ associated with the $T=0$ SG transition. As can be seen from Fig.19(a), significant finite-size effects appear at $\sigma=1$, which prevents us from reaching the asymptotic critical regime. Such an unsaturated behavior is also observed for the CG correlation length. These behaviors are somewhat in contrast the case of $\sigma=1.1$ shown in Fig.19(b), where an asymptotic critical behavior seems to be reached both in $\xi_{SG}$ and $\xi_{CG}$. In the latter cases, one can estimate the exponent $\nu_{SG}$ (or $\nu_{CG}$) associated with the $T=0$ transition from an asymptotic slope of the data.

%%
%\begin{figure}[!hbp]
%\begin{center}
%\includegraphics[scale=0.9]{Pqchi1.0.eps}
%\includegraphics[scale=0.9]{Pqdiag1.0.eps}
%\end{center}
%\caption{
%Overlap distribution function of the chirality (a) and of the spin (b) for $\sigma=1.0$ at a temperature $T=0.034$ below $T_{SG}$.
%}
%\end{figure}
%%

\subsection{$\sigma=1.1$}

\begin{figure}[!hbp]
\begin{center}
\includegraphics[scale=0.9]{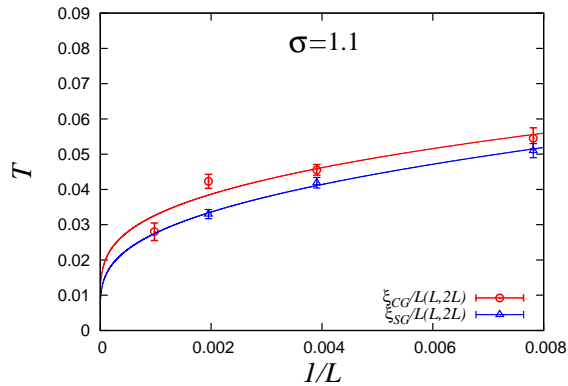}
\end{center}
\caption{
The (inverse) size dependence of the crossing temperatures of $\xi_{CG}/L$ and of $\xi_{SG}/L$ for $\sigma=1.1$.  Lines represent logarithmic-law fits of the form (19).
} 
\end{figure}

  We show in Fig.20 the size dependence of the crossing temperatures $T_{cross}(L)$ of $\xi_{CG}/L$ and of $\xi_{SG}/L$ for the case of $\sigma=1.1$. Here we find again that the power-law fit becomes unstable both for the SG and the CG, yielding indefinitely negative $T_{SG}$- or $T_{CG}$-value. We interpret this as suggesting the $T=0$ transition lying close to the upper critical $\sigma$ for both cases of the SG and the CG. As shown in Fig.19, the logarithmic plot as employed for $\sigma=1.0$ turns out to work pretty well also for both the SG and the CG at $\sigma=1.1$, with $\theta \simeq 2.8$ for the SG, and with $\theta \simeq 2.4$ for the CG.

 In our previous report of Ref.\cite{MatsuKawa07}, we indicated a nonzero $T_{CG}$ for $\sigma=1.1$ on the basis of $L\leq 1024$ data. In view of an intrinsic difficulty encountered near the upper-critical $\sigma$ in distinguishing a $T_{CG}=0$ transition with a large $\nu_{CG}$ from a $T_{CG}>0$ transition, however, the question of whether the CG transition persists at $\sigma=1.1$ is not completely clear.

\subsection{Phase diagram}

 The results obtained in the previous subsections are summarized in the $\sigma -T $ phase diagram of Fig.21. The spin-chirality decoupling occurs in the range $0.8\lsim \sigma \lsim 1.1$. By contrast, the standard spin-chirality coupling behavior with $T_{SG}=T_{CG} $ is realized for $\sigma \lsim 0.8$. The para-CG phase boundary might go beyond $\sigma=1$, touching the $T=0$ axis separately from the CG-SG phase boundary, although the possibility of it closing just at $\sigma=1$ simultaneously with the CG-SG phase boundary cannot be ruled out as shown by the thin dotted line in Fig.21.

\begin{figure}[!hbp]
\begin{center}
\includegraphics[scale=0.9]{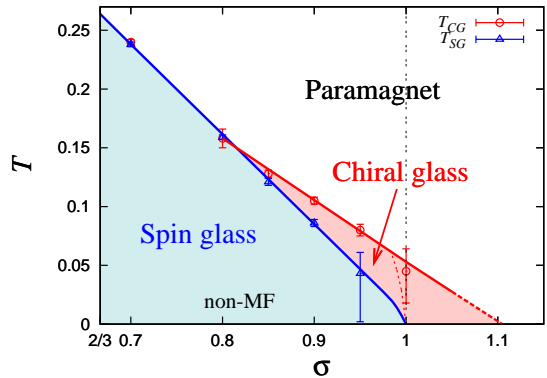}
\end{center}
\caption{
The range parameter $\sigma$ versus the temperature $T$ phase diagram of the 1D Heisenberg SG with a LR power-law interaction decaying with a distance $r$ $\propto r^{-\sigma}$. The red (blue) points are the chiral $T_{CG}$ (the spin $T_{SG}$) transition temperature. The spin-chirality decoupling occurs in the range $0.8 \lsim \sigma \lsim 1.1$, while more standard coupling behavior occurs in the range $\sigma \lsim 0.8$. The vertical dotted line represents the upper critical $\sigma$-value, $\sigma=1$. The lower critical $\sigma$-value is $\sigma=\frac{2}{3}$. The para-CG phase boundary might go beyond $\sigma=1$, touching the $T=0$ axis separately from the CG-SG phase boundary, although the possibility of it closing just at $\sigma=1$ simultaneously with the CG-SG phase boundary cannot be ruled out. 
}
\end{figure}

\section{Critical properties}

 In this section, we study the critical properties of the CG and the SG transitions on the basis of a finite-size scaling analysis of our data of the glass susceptibility and the correlation-length ratio. Below, we show detailed analyses for the cases of $\sigma=0.9$ and 0.8, each corresponding to the spin-chirality decoupling and coupling regimes, respectively. The critical properties for other $\sigma$-values are also studied, although for the $\sigma$-values other than $\sigma=0.9$ and 0.8 we quote only the resulting exponent values.

\subsection{$\sigma=0.9$}

 This value of $\sigma$ corresponds to the spin-chirality decoupling regime. From our analysis in the previous section, the CG and the SG transition temperatures were estimated to be $T_{CG}=0.105\pm 0.003$ and $T_{SG}=0.086\pm 0.003$, respectively. In our following analysis, we fix the $T_{CG}$- and the $T_{SG}$-values to these best values. The analysis in \S 4 suggested the presence of a significant correction-to-scaling term. The leading correction-to-scaling exponent was estimated to be $\theta_{CG}=\omega_{CG}+\frac{1}{\nu_{CG}}=1.2\pm 1.4$ and $\theta_{SG}=\omega_{SG}+\frac{1}{\nu_{SG}}=0.44\pm 0.07$.  In our analysis, we will include the effect of the correction-to-scaling  by fixing $\omega+\frac{1}{\nu}$ to these best values. We then estimate the two independent critical exponents characterizing the CG (SG) transitions, {\it i.e.\/}, the correlation-length exponent $\nu_{CG}$ ($\nu_{SG}$) and the critical-point-decay exponent $\eta_{CG}$ ($\eta_{SG}$).

We begin with the analysis of the critical properties of the CG transition. We employ the finite-size scaling forms for the correlation-length ratio $\xi_{CG}/L$ and for the CG susceptibility ${\chi}_{CG}$ given by eqs.(13) and (15), respectively. 

As shown in Fig.22, a reasonably good scaling is obtained both for the  CG correlation-length ratio $\xi_{CG}/L$ and for the CG susceptibility $\chi_{CG}$  by setting $\nu_{CG}=4.0$ and $\eta_{CG}=1.9$. The associated error bars are estimated  by examining by eyes the quality of the fit with varying the fitting parameters. We then get $\nu_{CG}=4.0\pm 0.3$ and $\eta_{CG}=1.9\pm 0.1$.

 We perform a similar finite-size scaling analysis also for the spin. We show in Fig.23  the finite-size-scaling plots of the SG correlation-length ratio $\xi_{SG}/L$ and of the SG susceptibility ${\chi}_{SG}$ with the correction term. Both quantities can be scaled reasonably well by setting $\nu_{SG}=3.3$ and $\eta_{SG}=1.2$.  Our final estimates are then $\nu_{SG}=3.3\pm 0.3$ and $\eta_{SG}=1.2\pm 0.1$. Note that the obtained value of $\eta_{SG}$ is fully consistent with the analytical expression obtained for the LR case $\eta_{SG}=3-2\sigma$.

\begin{figure}[ht]
\begin{center}
\includegraphics[scale=0.9]{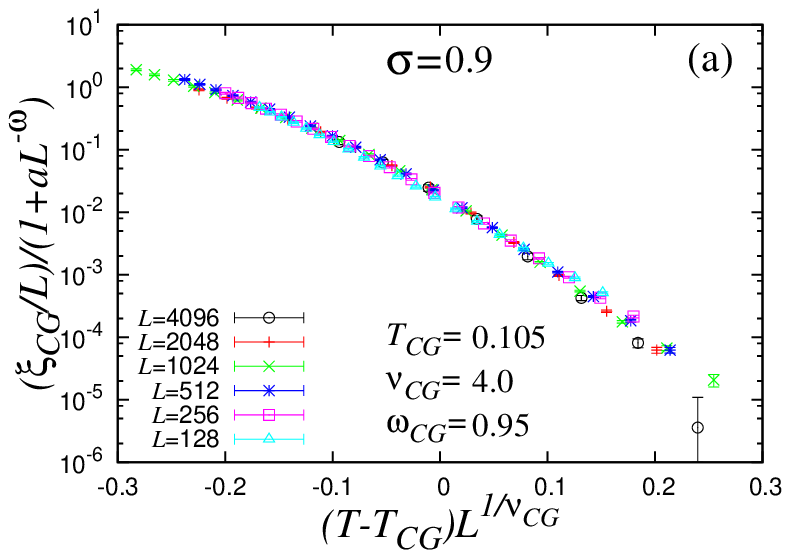}
\includegraphics[scale=0.9]{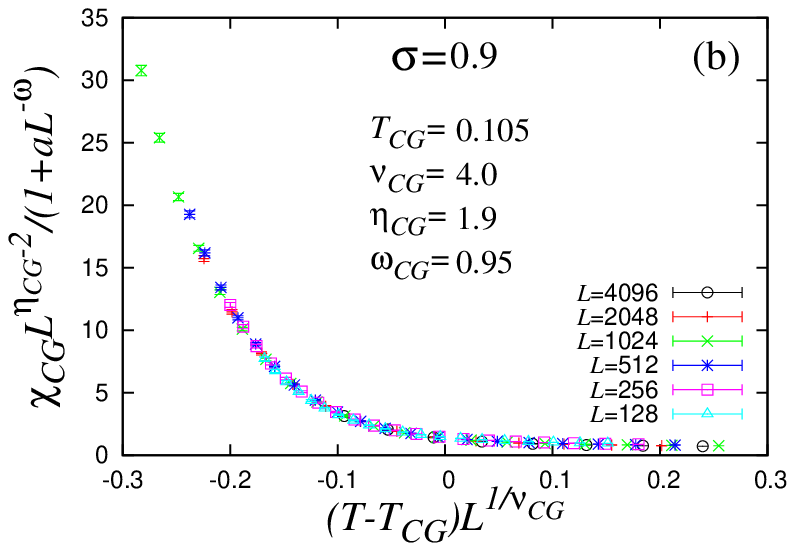}
\end{center}
\caption{
(Color online) Finite-size-scaling plots of the CG correlation-length ratio $\xi_{CG}/L$ (a), and of the CG susceptibility $\chi_{CG}$ (b), for the case of $\sigma=0.9$, where the correction-to-scaling effect is taken into account. The CG transition temperature and the leading correction-to-scaling exponents are fixed to $T_{CG}=0.105$ and $\omega_{CG}+\frac{1}{\nu_{CG}}=1.2$ as determined in \S 4. The best data collapse for $\xi_{CG}/L$ is obtained with $\nu_{CG}=4.0$, while that for $\chi_{CG}$ is obtained with $\nu_{CG}=4.0$ and $\eta_{CG}=1.9$.
}
\end{figure}
\begin{figure}[ht]
\begin{center}
\includegraphics[scale=0.9]{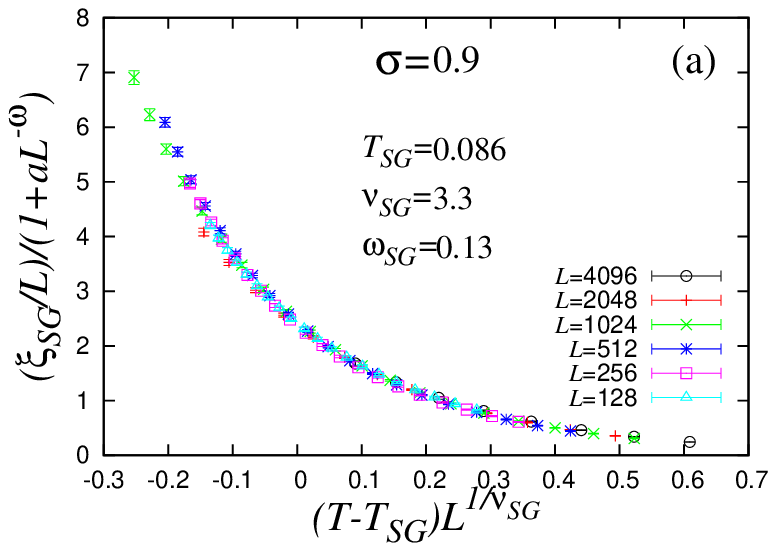}
\includegraphics[scale=0.9]{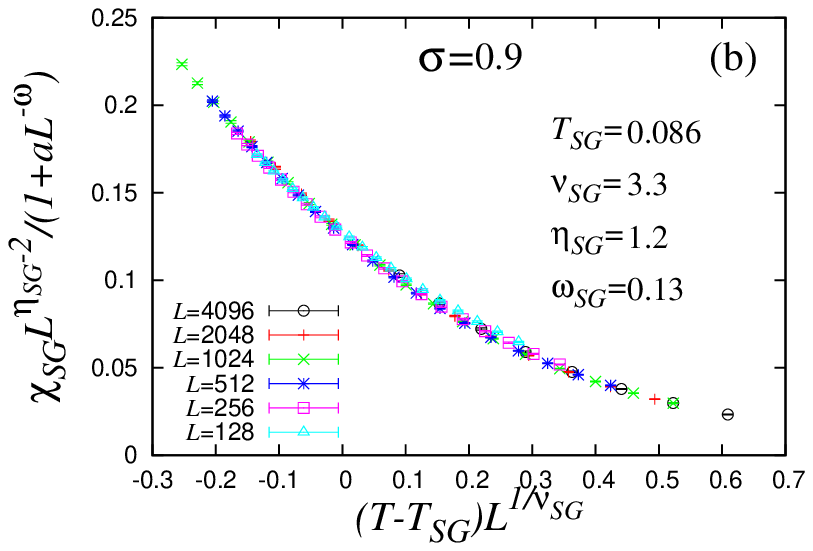}
\end{center}
\caption{
(Color online) Finite-size-scaling plots of the SG correlation-length ratio $\xi_{SG}/L$ (a), and of the SG susceptibility $\chi_{SG}$ (b), for the case of $\sigma=0.9$, where the correction-to-scaling effect is taken into account. The SG transition temperature and the leading correction-to-scaling exponents are fixed to $T_{SG}=0.086$ and $\omega_{SG}+\frac{1}{\nu_{SG}}=0.44$ as determined in \S 4. The best data collapse for $\xi_{SG}/L$ is obtained with $\nu_{SG}=3.3$, while that for $\chi_{SG}$ is obtained with $\nu_{SG}=3.3$ and $\eta_{SG}=1.2$.
}
\end{figure}

\subsection{$\sigma=0.8$}

 This value of $\sigma$ corresponds to the spin-chirality coupling regime. From our analysis in the previous section, the CG and the SG transition temperatures were estimated to be $T_{CG}=0.158\pm 0.008$ and $T_{SG}=0.159\pm 0.002$, while the leading correction-to-scaling exponent to be $\theta_{CG}=\omega_{CG}+\frac{1}{\nu_{CG}}=0.62\pm 1.95$ and $\theta_{SG}=\omega_{SG}+\frac{1}{\nu_{SG}}=0.47\pm 0.11$.

Via the finite-size-scaling analysis shown in Figs. 24 and 25, the CG exponents are determined to be $\nu_{CG}=4.0\pm 0.5$ and $\eta_{CG}=2.0\pm 0.1$, while the SG exponents are determined to be $\nu_{SG}=3.7\pm 0.3$ and $\eta_{SG}=1.4\pm 0.1$. Thus, we get $\nu_{CG}=\nu_{SG}$ within the error bar, which is consistent with the expected spin-chirality coupling behavior, {\it i.e.\/}, only one diverging length scale at the transition. Again, the obtained value of $\eta_{SG}$ is fully consistent with the analytically obtained expression $\eta_{SG}=3-2\sigma=1.4$. Note that $\eta_{CG}$ and $\eta_{SG}$ need not be equal even in the spin-chirality coupling case, the spin and the chirality carrying their own anomalous dimensions but with a common diverging length scale.

\begin{figure}[ht]
\begin{center}
\includegraphics[scale=0.9]{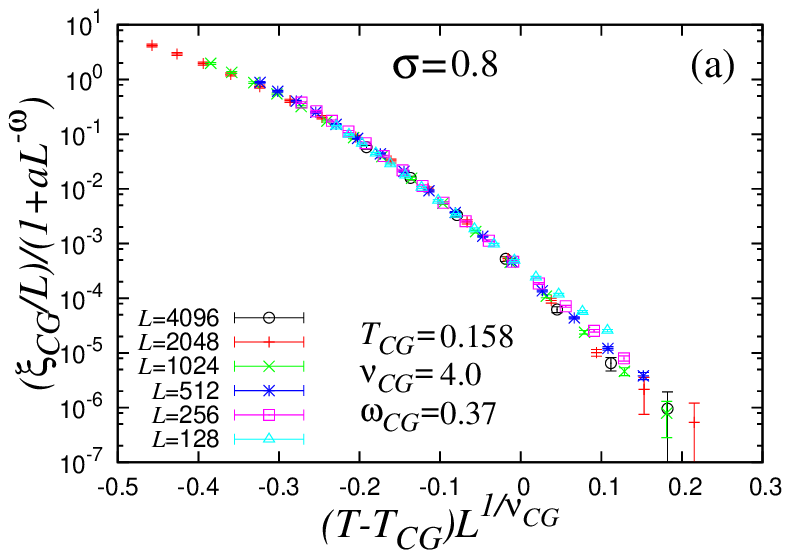}
\includegraphics[scale=0.9]{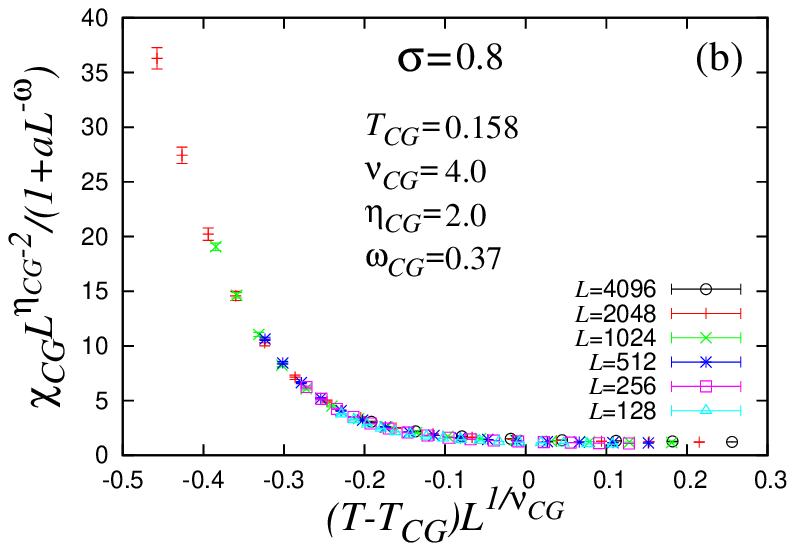}
\end{center}
\caption{
(Color online) Finite-size-scaling plots of the CG correlation-length ratio $\xi_{CG}/L$ (a), and of the CG susceptibility $\chi_{CG}$ (b), for the case of $\sigma=0.8$, where the correction-to-scaling effect is taken into account. The CG transition temperature and the leading correction-to-scaling exponents are fixed to $T_{CG}=0.158$ and $\omega+\frac{1}{\nu}=0.62$ as determined in \S 4. The best data collapse for $\xi_{CG}/L$ is obtained with $\nu_{CG}=4.0$, while that for $\chi_{CG}$ is obtained with $\nu_{CG}=4.0$ and $\eta_{CG}=2.0$.
}
\end{figure}
\begin{figure}[ht]
\begin{center}
\includegraphics[scale=0.9]{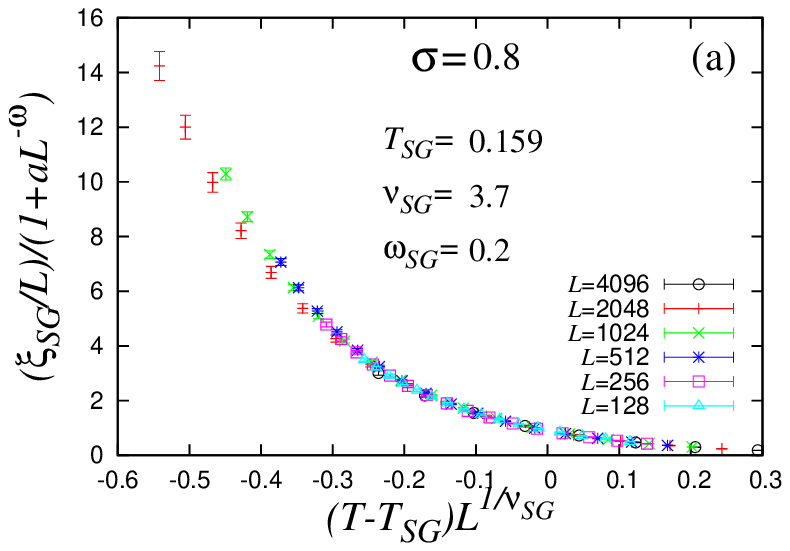}
\includegraphics[scale=0.9]{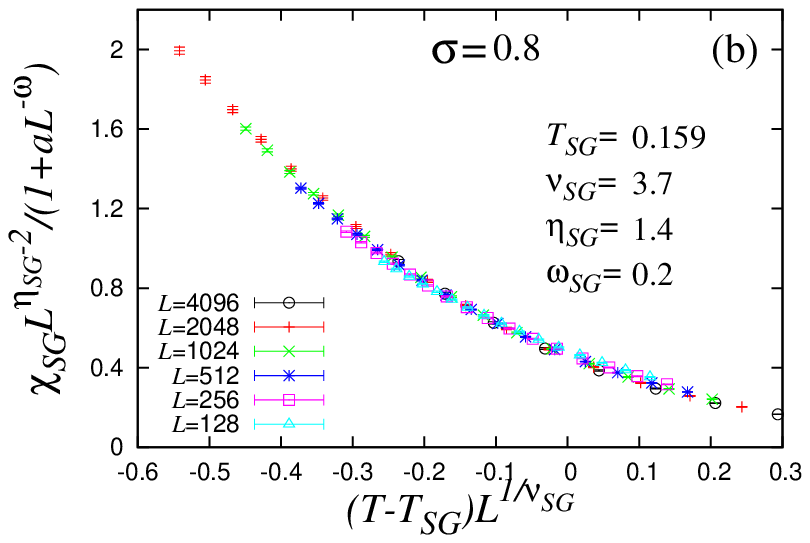}
\end{center}
\caption{
(Color online) Finite-size-scaling plots of the SG correlation-length ratio $\xi_{SG}/L$ (a), and of the SG susceptibility $\chi_{SG}$ (b), for the case of $\sigma=0.8$, where the correction-to-scaling effect is taken into account. The SG transition temperature and the leading correction-to-scaling exponents are fixed to $T_{SG}=0.159$ and $\omega_{SG}+\frac{1}{\nu_{SG}}=0.47$ as determined in \S 4. The best data collapse for $\xi_{SG}/L$ is obtained with $\nu_{SG}=3.7$, while that for $\chi_{SG}$ is obtained with $\nu_{SG}=3.7$ and $\eta_{SG}=1.4$.
}
\end{figure}

\subsection{Other values of $\sigma$}

 We have performed similar finite-size-scaling analyses for other values of $\sigma$, and the resulting exponents $\nu_{CG}$, $\eta_{CG}$, $\nu_{SG}$ and $\eta_{SG}$ are summarized in Fig.26. The analytically obtained $\eta_{SG}$-value is also included in the figure. As demonstrated in Fig.19(a), finite-size effects are so severe at $\sigma=1$ that we cannot give a reliable estimate of $\nu_{SG}$ at $\sigma=1$. Nevertheless, as can be deduced from the logarithmic fit made in Fig.18 and from the non-convergent size dependence observed in Fig.19(a), the $\nu_{SG}$-value at $\sigma=1$ could be quite large in the thermodynamic limit, which is not inconsistent with $\nu_{SG}=\infty$ generically expected at the upper-critical $\sigma$.

 Several points are to be noticed here. (i) The estimated $\eta_{SG}$ agrees well with the analytical expression $\eta_{SG}=3-2\sigma$ over the entire $\sigma\leq 1$ regime. (ii) The estimated $\eta_{CG}$ is greater than $\eta_{SG}$ over the entire $\sigma$ range studied.  (iii) At $\sigma=2/3$, $\nu_{SG}$ is expected to approach the MF value $\nu_{SG}=3$. The $\nu_{SG}$-value obtained here toward $\sigma=2/3$ appears to be somewhat greater than this value. Presumably, a logarithmic correction expected at the lower-critical $\sigma=2/3$ might make an accurate estimate of the exponent difficult around $\sigma=2/3$, and the observed exponent might be an effective exponent. Meanwhile, our present estimate of $\nu_{SG}$ at $\sigma=0.7$, $\nu_{SG}=4.0\pm 0.5$, is close to the ones obtained for the 1D Ising SG with a LR power-law interaction at $\sigma=0.69$, {\it i.e.\/}, $\nu=3.7\pm 0.6$ \cite{BhattYoung86} or $3.8\pm 0.4$ \cite{Leuzzi99}, and the one at $\sigma=0.75$, {\it i.e.\/}, $\nu=4.5\pm 0.2$ \cite{Leuzzi99}, $3.3\pm 0.4$ \cite{Katzgraber03} or $4.0\pm 0.5$ \cite{Leuzzi08}.

\begin{figure}[ht]
\begin{center}
\includegraphics[scale=0.9]{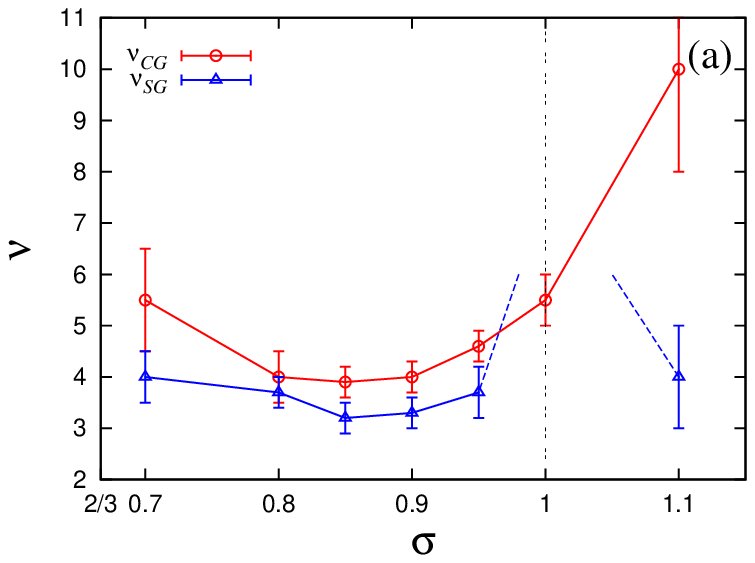}
\includegraphics[scale=0.9]{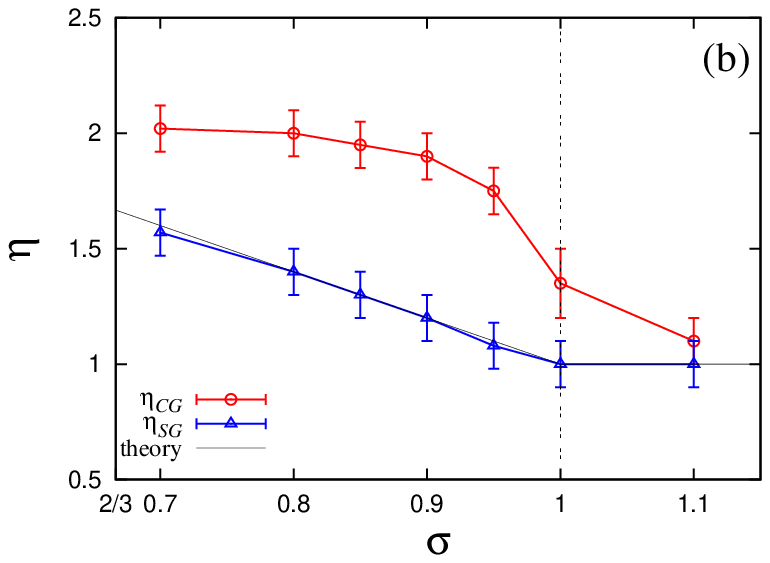}
\end{center}
\caption{
(Color online) The $\sigma$-dependence of the correlation-length exponents $\nu_{CG}$ and  $\nu_{SG}$ (a), and the anomalous-dimension exponents $\eta_{CG}$ and $\eta_{SG}$ (b). The analytically obtained $\eta_{SG}$-values, {\it i.e.\/}, $\eta_{SG}=3-2\sigma$ for $\sigma \leq 1$ and $\eta_{SG}=1$ for $\sigma\geq 1$, are also indicated by lines.
}
\end{figure}
\section{Summary and discussion}

 We performed a large-scale equilibrium MC simulation on the 1D Heisenberg SG with LR power-law interactions, paying attention to the SG and the CG orderings and the possible spin-chirality decoupling phenomena of the model. This one-dimensional SG model might have an advantage over the 3D model that larger linear sizes can be studied. Furthermore, by continuously varying and even fine-tuning the power-law exponent $\sigma$, which plays a role of effective ``dimensionality'', different types of ordering behaviors are realized.

 By calculating various physical quantities including the correlation-length ratio, the susceptibility ratio, the Binder ratio, and the overlap distribution function up to the sizes as large as $L=4096$ for various $\sigma$-values in a range of $0.7 \leq \sigma \leq 1.1$, we obtained a strong numerical evidence for the occurrence of the spin-chirality decoupling behavior in the range $0.8\lsim \sigma \lsim 1.1$, while that of the standard spin-chirality coupling behavior with $T_{SG}=T_{CG} $ in the range  $\sigma \lsim 0.8$. Our results are summarized in the temperature-$\sigma$ phase diagram of Fig.21.

 Even in the spin-chirality decoupling regime $0.8\lsim \sigma \lsim 1.1$,  the spin and the chirality often behave in a similar way on shorter length scale of $L\lsim 500\sim 1000$, while  on longer length scale the chirality shows a much stronger ordering tendency than the spin. The observation supports the view of the spin-chirality coupling behavior at shorter length scale crossing over to the spin-chirality decoupling behavior at longer length scale \cite{Kawamura07,Kawamura09}

 The Binder ratio exhibits a weak one-step-like RSB feature in the CG ordered state, at least in the $\sigma$-range where the model exhibits the spin-chirality decoupling behavior. It should be noticed, however, that the one-step-like RSB feature observed in the present 1D LR model is much weaker than the one observed in the CG ordered state of the 3D Heisenberg SG model: The dip of the chiral Binder ratio is very shallow and the central peak of the chiral overlap distribution function is hardly discernible. By contrast, for $\sigma \lsim 0.8$ where the model exhibits the standard spin-chirality coupling behavior, we observed a full RSB feature consistently with the MF picture \cite{ImaKawa03}. 

 We also studied the critical properties of the model based on a finite-size-scaling analysis. The resulting exponent values are summarized in Fig.26. The behavior of the SG exponent $\eta_{SG}$ is consistent with the analytical result obtained from the RG analysis. 

 We try to further examine the possible $d$-$\sigma$ correspondence.  The behavior of $\xi_{CG}/L$ of the 3D short-range model looks similar to those of $\sigma=0.9$ or $\sigma=0.95$ of the 1D LR model. To our knowledgeable, no data of the CG and SG correlation length ratios are available for higher dimensional Heisenberg SG, which prevents us from making a further comparison. Another point to be noticed is that the increasing/decreasing tendency of the size-dependence of $T_{cross}(L)$ of $\xi_{CG}/L$ changes from the decreasing (with $L$) behavior for $\sigma=1.0$ and 0.95 to the nearly constant behavior for $\sigma =0.9$ and 0.85, and then to the increasing behavior for $\sigma=0.8$, and 0.7. In comparison with those of the 3D SR models, it appears  that the 3D SR Heisenberg SG corresponds to a $\sigma$-value somewhere between $\sigma=0.9$ or $\sigma=0.95$. Meanwhile, the 2D SR Heisenberg SG corresponds to a $\sigma$-value around $\sigma=1.1$.

 Overall, the ordering behavior of the Heisenberg SG can roughly be classified into three regimes, {\it i.e.\/}, the spin-chirality coupling behavior for smaller $\sigma$ (larger $d$), the spin-chirality decoupling behavior for intermediate $\sigma$ (intermediate $d$), and the zero-temperature transition behavior for larger $\sigma$ (smaller $d$). Thus, our present study on the 1D LR Heisenberg SG serves to provide an overall picture of the ordering behavior of the Heisenberg SG from a wider perspective.

  This study was supported by Grant-in-Aid for Scientific Research on Priority Areas ``Novel States of Matter Induced by Frustration'' (19052006). Numerical calculation was performed at ISSP, Tokyo University, and at YITP, Kyoto University. The authors are thankful to I.A. Campbell and H. Yoshino for useful discussion and suggestion.

\end{document}